\documentclass[12pt, reqno]{amsart}

\usepackage{a4}
\usepackage{amsmath}
\usepackage{diagrams}               
\usepackage{amssymb}

\usepackage{graphicx}               

 \theoremstyle{plain}            
 \newtheorem{theorem}{Theorem}[section]
 \newtheorem*{maintheorem*}{Main Theorem}
 \newtheorem*{maintheorem.}{Main Theorem~1'}
 \newtheorem{maintheorem}{Main Theorem}
 \newtheorem{proposition}[theorem]{Proposition}
 \newtheorem{lemma}[theorem]{Lemma}

 \theoremstyle{definition}       
 \newtheorem{definition}[theorem]{Definition}

 \theoremstyle{remark}
 \newtheorem{remark}[theorem]{Remark}
 \newtheorem{example}[theorem]{Example}
 \newtheorem{consequence}{Consequence}
 \numberwithin{equation}{section}

\DeclareMathOperator{\tr}     {tr}
\DeclareMathOperator{\dist}   {dist}
\DeclareMathOperator{\dom}    {dom}

\DeclareMathOperator{\spec}   {spec}
\DeclareMathOperator{\supp}   {supp}
\DeclareMathOperator{\vol}    {vol}

\DeclareMathOperator{\intr} {int} 

\newlength{\maxbreite}%
\newlength{\maxhoehe}%
\newlength{\maxtiefe}%

\newcommand{\stelldrueber}[3][0pt]{
  \settowidth{\maxbreite}{#3}%
  \settoheight{\maxhoehe}{#3}%
  \settodepth{\maxtiefe}{#2}%
  \addtolength{\maxhoehe}{\maxtiefe}%
  {\makebox[\maxbreite]{\raisebox{\maxhoehe}{\hspace{#1}#2}}%
  \makebox[0pt][r]{#3}}%
}

\newcommand{\overcirc}[1]       
{\stelldrueber[.45ex]{$\scriptscriptstyle\circ$}{${#1}$}}

\newcommand{\R}{\mathbb{R}} 
\newcommand{\C}{\mathbb{C}} 
\newcommand{\N}{\mathbb{N}} 
\newcommand{\Z}{\mathbb{Z}} 
\newcommand{\Sphere}{\mathbb{S}} 
\newcommand{\Torus}{\mathbb{T}} 
\newcommand{\Hyp}{\mathbb{H}} 
\newcommand{\eps}{\varepsilon} 
\renewcommand{\phi}{\varphi}   
\newcommand{\HS}{\mathcal H} 

\newcommand{\End} [1]{\mathcal L({#1})}     
\newcommand{\Contsymb}{\mathsf C}           
\newcommand{\Ci} [1]{\Contsymb^\infty ({#1})}       
\newcommand{\Cci}[1]{\Contsymb_{\mathrm c}^\infty ({#1})} 

\newcommand{\Lsymb}{\mathsf L}              
\newcommand{\Lsqr}[1]{\Lsymb_2({#1})}            
\newcommand{\lsqr}[1]{\ell_2({#1})}         
\newcommand{\Linfty}[1]{\Lsymb_\infty({#1})}     

\newcommand{\Sobsymb}{\mathsf H}           
\newcommand{\Sob}[2][1]{\Sobsymb^{#1}({#2})}
\newcommand{\Sobn}[2][1]{{\Sobsymb_\circ^{#1}({#2})}}

\newcommand{\norm}[2][{}]{\|{#2}\|_{#1}}    
\newcommand{\normsqr}[2][{}]{\|{#2}\|^2_{#1}} 
\newcommand{\iprod}[3][{}]{\langle{#2},{#3}\rangle_{#1}}  

\newcommand{\bd}  {\partial}                

\newcommand{\restr}[1]{{\restriction}_{#1}} 
\newcommand{\map}[3]{{#1}\colon{#2}\longrightarrow{#3}} 
\newcommand{\set}[2]{\{ \, #1 \, | \, #2 \, \} } 

\usepackage{bbm}         

\newcommand{\1}{\mathbbm 1}                    

\newcommand{\Neu}{{-}}              
\newcommand{\Dir}{{+}}              
\newcommand{\DirNeu}{\pm}

\newcommand{\laplacian}[1]{\Delta_{{#1}}}       
\newcommand{\laplacianD}[1]{\Delta^\Dir_{{#1}}} 
\newcommand{\laplacianDN}[1]{\Delta^{\DirNeu} _{{#1}}}
\newcommand{\laplacianN}[1]{\Delta^\Neu_{{#1}}} 

\newcommand{\CiEq}[1]{\Contsymb^\infty_{\mathrm {eq}} ({{#1}})} 
\newcommand{\CiR}[1]{\Contsymb^\infty_\rho({{#1}})}   
\newcommand{\SobR}[2][1]{\Sobsymb^{#1}_\rho({{#2}})}

\newcommand{\laplacianR}[1]{\Delta^\rho_{{#1}}} 
\newcommand{\laplacianZ}[1]{\Delta^\mathrm{eq}_{{#1}}(z)} 

\newcommand{\EW}[1]{\lambda_{#1}}        
\newcommand{\EWD}[1]{\lambda^\Dir_{#1}}  
\newcommand{\EWN}[1]{\lambda^\Neu_{#1}}  
\newcommand{\EWDN}[1]{\lambda^{\DirNeu}_{#1}}
\newcommand{\EWR}[1]{\lambda^\rho_{#1}}  

\newcommand{\Dint}[4][{}]{\int_{{#2}}^{{#1}} #3 \, \mathrm d #4} 
\newcommand{\Oint}[3]{\int_{{#1}}^\oplus #2 \, \mathrm d #3} 
 
\newcommand{\hG}{\widehat \Gamma}         
\newcommand{\tg}{{\widetilde \gamma}}         
\newcommand{\Unitary}[1]{\mathcal U(#1)}    

\newcommand{\OintZ}[1]{\Oint Z {#1} z}          

\newcommand{\al}[1]{\mathcal{#1}}

\begin{document}

\title[Existence of spectral gaps and residually finite groups]
{Existence of spectral gaps, covering manifolds and residually
  finite groups}
 \author{Fernando Lled\'o}
 \address{Institut f\"ur Reine und Angewandte Mathematik,
        Rheinisch-Westf\"alische Technische Hochschule Aachen,
        Templergraben 55,
        D-52062 Aachen,
        Germany (on leave)\\
     Department of Mathematics, University Carlos~III Madrid,
     Avda.~de la Universidad~30, E-28911 Leganes (Madrid),
     Spain }
  \email{lledo@iram.rwth-aachen.de {\em and} flledo@math.uc3m.es}
  \author{Olaf Post}
\address{Institut f\"ur Mathematik,
         Humboldt-Universit\"at zu Berlin,
         Rudower Chaussee~25,
         D-12489 Berlin,
         Germany}
\email{post@math.hu-berlin.de}
\date{\today}
\dedicatory{Dedicated to Volker En{\ss} on his 65th birthday}



\begin{abstract}
  In the present paper we consider Riemannian coverings $(X,g) \to
  (M,g)$ with residually finite covering group $\Gamma$ and compact
  base space $(M,g)$. In particular, we give two general procedures
  resulting in a family of deformed coverings $(X,g_\eps) \to
  (M,g_\eps)$ such that the spectrum of the Laplacian
  $\laplacian{(X_\eps,g_\eps)}$ has at least a prescribed finite
  number of spectral gaps provided $\eps$ is small enough.
    
  \sloppy 
  If $\Gamma$ has a positive Kadison constant, then we can
  apply results by Br\"uning and Sunada to deduce that $\spec
  \laplacian{(X,g_\eps)}$ has, in addition, band-structure and there
  is an asymptotic estimate for the number $\mathcal N(\lambda)$ of
  components of $\spec {\laplacian {(X,g_\eps)}}$ that intersect the
  interval $[0,\lambda]$. We also present several classes of
  examples of residually finite groups that fit with our construction
  and study their interrelations. Finally, we mention several possible
  applications for our results.
\end{abstract}

\maketitle


%
\section{Introduction}
%
\label{sec:intro}

Spectral properties of the Laplacian on a compact manifold is a
well-established and still active field of research. Much less is
known on the spectrum of \emph{non-compact} manifolds. We restrict
ourselves here to the class of non-compact \emph{covering} manifolds
$X \to M$ with compact quotient $M$, in which the covering group
$\Gamma$ plays an important role. In the open problem section of
\cite[Ch.~IX, Problem~37]{schoen-yau:94}, Yau posed the question about
the nature and the stability of the (purely essential) spectrum of
such a covering $X \to M$.

The aim of this paper is to provide a large class of examples of
Riemannian coverings $X \to M$ having spectral gaps in the essential
spectrum of its Laplacian~$\laplacian X$. Here, a spectral gap is a
non-void open interval $(\alpha, \beta)$ with $(\alpha, \beta) \cap
\spec {\laplacian X } = \emptyset$ and $\alpha, \beta \in \spec
{\laplacian X}$.  The manifolds $X$ and $M$ are $d$-dimensional,
$d\geq 2$, and we denote by $D$ a fundamental domain associated to
this covering.  The main idea for producing spectral gaps is to
construct a family of Riemannian metrics $(g_\eps)_{\eps>0}$ on $X$
such that the length scale w.r.t.\ the metric $g_\eps$ is of order
$\eps$ at the boundary of a fundamental domain $D$ and unchanged
elsewhere (cf.~Figure~\ref{fig:per-mfd}). If such a fundamental domain
exists, we say that the family of metrics $(g_\eps)$ \emph{decouples}
the manifold $X$. The covering $X \to M$ with a decoupling family of
metrics $(g_\eps)$ ``converges'' in a sense to be specified below to a
limit covering consisting of the infinite disjoint (``decoupled'') union of the
limit quotient manifold $N$ which are again $d$-dimensional (see
Subsection~\ref{ssec:outline} and Section~\ref{sec:construye} for
details). We stress that the curvature does not remain bounded as
$\eps \to 0$; in contrast to degeneration of Riemannian metrics under
curvature bounds developed~e.g.\ in~\cite{cheeger:01}.  All groups
$\Gamma$ are assumed to be discrete and finitely generated throughout
the present article.

\subsection{Statement of the main results}
\begin{maintheorem}[cf.~Theorem~\ref{thm:gaps.res.fin}]
  \label{mthm:1}
  Suppose that $X \to M$ is a Riemannian covering with residually
  finite covering group $\Gamma$ and metric $g$.  Then by a local 
  deformation of $g$ we
  construct a family of metrics $(g_\eps)$ decoupling $X$, such
  that for each $n \in \N$ there exists $\eps_n>0$ where $\spec
  {\laplacian {(X,g_{\eps_n})}}$ has at least $n$ gaps, i.e.\ $n+1$
  components as subset of $[0,\infty)$.
\end{maintheorem}
Basically, we will give two different constructions for the family of
manifolds $(X,g_\eps)$: first, ``adding small handles'' to a given
manifold $(N,g)$ and second, a conformal perturbation of $g$. As a
set, $(X,g_\eps)$ converges to a limit manifold consisting of
infinitely many disjoint copies of the limit quotient manifold $N$ as
$\eps \to 0$.

A \emph{residually finite} group is a countable
discrete group such that the intersection of all its normal subgroups
of finite index is trivial.  Roughly speaking, a residually
  finite group has many normal subgroups of finite index.
Geometrically, a covering with a residually finite covering group can
be approximated by a sequence of finite coverings $M_i \to M$ (a
\emph{tower of coverings}). The class of residually finite groups is
very large, containing e.g.~finitely generated abelian groups, type~I
groups (i.e.\ finite extensions of $\Z^r$), free groups or finitely
generated subgroups of the isometries of the $d$-dimensional
hyperbolic space $\Hyp^d$ (cf.\ Section~\ref{sec:res.fin}).

Denote by $\mathcal N(g,\lambda)$ the number of components of
$\spec\laplacian {(X,g)}$ which intersect the interval $[0, \lambda]$.
Our result gives a \emph{lower} bound on $\mathcal N(g,\lambda)$, in
particular, we can reformulate the Main~Theorem~\ref{mthm:1} as
follows: \emph{For each $n \in \N$ there exists
  $g=g_{\eps_n}$ such that $\mathcal N(g,\lambda) \ge n+1$.}

Using the Weyl eigenvalue asymptotic on the limit $d$-dimensional
manifold $(N,g)$ associated to the decoupling family $(g_\eps)$ on $X
\to M$, we obtain the following asymptotic lower bound on the number
of gaps (where $\omega_d$ denotes the volume of the $d$-dimensional
Euclidean unit ball):
\begin{maintheorem}[cf.~Theorem~\ref{thm:band}]
  Assume that the covering group is residually finite and that the
  spectrum of the Laplacian on the limit manifold $(N,g)$ is simple,
  i.e.~ all eigenvalues have multiplicity one.  Then for each $\lambda
  \ge 0$ there exists $\eps(\lambda)>0$ such that
  \begin{equation*}
    \liminf_{\lambda\to\infty} 
       \frac{\mathcal N (g_{\eps(\lambda)}, \lambda)}
       {(2\pi)^{-d} \omega_d \vol (N,g) \lambda^{d/2}}
    \ge 1.
  \end{equation*}
\end{maintheorem}
The assumption on the spectrum of $(N,g)$ is natural since $\mathcal
N(g,\lambda)$ counts components in the spectrum \emph{without}
multiplicity.

A priori, the number of gaps $\mathcal N(g,\lambda)$ could be
infinite, e.g.~if $\spec {\laplacian {(X,g)}}$ contains a Cantor set.
But Br\"uning and Sunada showed in~\cite{bruening-sunada:92} that for
covering groups $\Gamma$ with positive \emph{Kadison constant}
$C(\Gamma)>0$ (cf.~Section~\ref{sec:kadison}) asymptotic upper bound
\begin{equation*}
   \limsup_{\lambda\to\infty} \frac{\mathcal N(g, \lambda)} 
    {(2\pi)^{-d} \omega_d \vol (M,g) \lambda^{d/2}}
    \le \frac 1 {C(\Gamma)}
\end{equation*}
holds.  In particular, $\mathcal N(g,\lambda)$ is finite, and the
spectrum of $\laplacian {(X,g)}$ does not contain Cantor-like subsets.
Applying these results to our situation we give a partial answer on
the question of Yau of the nature of the spectrum:
\begin{maintheorem}[cf.~Theorem~\ref{thm:band}]
  \sloppy Suppose that $X \to M$ is a Riemannian $\Gamma$-covering
  with decoupling family of metrics $(g_\eps)$, where $\Gamma$ is a
  residually finite group that has positive Kadison constant
  $C(\Gamma)>0$. Then $\spec \laplacian {(X,g_\eps)}$ has
  band-structure, i.e.~$\mathcal N(g_\eps,\lambda)~<~\infty$ for all
  $\lambda \ge 0$ and $\mathcal N(g_\eps,\lambda)$ can be made
  arbitrary large provided $\eps$ is small and $\lambda$ is large
  enough.
\end{maintheorem}
Some examples of groups with positive Kadison
constant and which are residually finite are finitely generated,
abelian groups, the free (non-abelian) group in $r \ge 2$ generators
or fundamental groups of compact, orientable surfaces (see also
Section~\ref{sec:examples}).

\subsection{Motivation and related work}

A main motivation for our work comes from the spectral theory of
Schr\"odinger operators $H=-\Delta+V$ on $\R^d$, $d \ge 2$, with $V$
periodic w.r.t.~the action of a discrete abelian
group~$\Gamma_{\mathrm{ab}}=\Z^d$ on $\R^d$. For such operators, it is
a well known fact that if $V$ has high barriers near the boundary of a
fundamental domain $D$, then gaps appear in the spectrum of $H$. In
this way, the potential $V$ essentially decouples the fundamental
domain $D$ from its neighbouring domains (see \cite{hempel-post:03}
for an overview on this subject). 

A natural generalisation into a geometric context is to replace the
periodic structure $(\R^d, \Z^d)$ by a Riemannian covering $X \to M$
with a discrete (in general non-abelian) group $\Gamma$. Our work
shows that the decoupling effect of the potential $V$ can be replaced
purely by geometry, in particular by the decoupling family of metrics
$(g_\eps)$ on $X \to M$. From a quantum mechanical or probabilistic
point of view, the correspondence seems to be natural: One has a small
probability to find a particle (with low energy) in a region with a
high potential barrier or where the manifold $(X,g_\eps)$ is very thin
and the absolute value of the curvature is very large.

It was already observed by, e.g., Br\"uning, Gruber, Kobayashi, Ono
and Sunada \cite{bruening-sunada:92,gruber:01, sunada:90,kos:89} that
many properties of the spectrum of a periodic Schr\"odinger operator
(e.g.~band-structure, Bloch's property etc.)  generalise to the
context of Riemannian coverings.  An important difference is the
existence of $\Lsymb_2$-eigenvalues in the context of manifolds
(cf.~\cite{kos:89}).  Such eigenvalues cannot occur in the spectrum of
a periodic Schr\"odinger operator on $\R^d$ (cf.~\cite{sunada:90}).

The existence of (covering) manifolds with spectral gaps has also been
established by Br\"uning, Exner, Geyler and Lobanov
in~\cite{beg:03,bgl:05}.  They couple compact manifolds by points or
line-segments with certain boundary condition at the coupling points;
the point coupling corresponds to the case $\eps=0$ in our situation
(with decoupled boundary condition). The case of abelian \emph{smooth}
coverings has been established in \cite{post:03a} (cf.~also the
references therein). Spectral gaps of Schr\"odinger operators on the
hyperbolic space have been analysed in~\cite{karp-peyerimhoff:00}. For
other manifolds with spectral gaps (not necessarily periodic), we
refer to \cite{exner-post:05, post:06}. Under certain topological
restrictions on the middle degree homology group one can show the
existence of spectral gaps also for the differential form Laplacian on
a $\Z$-covering (see~\cite{acp:pre07}).

Some further interesting results on the group $\Gamma$ and spectral
properties of a Riemannian $\Gamma$-covering were shown by
Brooks~\cite{brooks:81}, e.g.\ that $\Gamma$ is amenable iff $0 \in
\spec \laplacian X$.  Moreover, Brooks~\cite{brooks:86} provided a
combinatorial criterion whether the second eigenvalue of $\laplacian
{M_i}$ is bounded from below as $i \to \infty$, where $M_i \to M$ is a
tower of coverings.

For physical applications of our results we refer to
Section~\ref{sec:outlook}.  Let us finish with two consequences of our
result giving partial answers to the question of Yau on the nature and
stability of the spectrum of $\laplacian X$:

\begin{consequence}[Manifold with given spectrum]
  First, we can solve the following inverse spectral problem: Given a
  compact (connected) manifold $N$ of dimension $d \ge 3$ and a
  sequence of numbers $0=\lambda_1(0)<\ldots<\lambda_n(0)$ it is
  possible to construct a metric $g$ on $N$ having exactly the numbers
  $\lambda_k(0)$ as first $n$ eigenvalues with multiplicity $1$
  (cf.~\cite{colin:87}).  Then, applying our Main Theorem~3 and using
  the relation between $\spec \laplacian {(X,g_\eps)}$ and $\spec
  \laplacian {(N,g)}$ we can construct a covering $X \to M$ with
  decoupling family $(g_\eps)$ having band spectrum close to the given
  points $\{\lambda_k(0)\}$, $k=1,\dots, n$. The covering $(X,g_\eps)
  \to (M,g_\eps)$ is obtained roughly by joining copies of $N$ through
  small, thin cylinders (see first construction mentioned below). In
  particular, we have constructed a covering manifold with
  approximatively given spectrum in a finite spectral interval
  $[0,\lambda]$, \emph{independently} of the covering group!
\end{consequence}

\begin{consequence}[Instability of gaps]
  Suppose $X=\Hyp^d$ is the $d$-dimensional ($d \ge 3$) hyperbolic
  space (or more generally, a simply connected, complete, symmetric
  space of non-compact type) with its natural metric $g$. It is known,
  that $\laplacian {(X,g)}$ has no spectral gaps, in particular $\spec
  {\laplacian {(X,g)}} = [\lambda_0, \infty)$ for some constant
  $\lambda_0 \ge 0$ (see e.g.~\cite{donnelly:79}). Let $\Gamma$ be a
  finitely generated subgroup of the isometries of $X$ such that $M =
  X/\Gamma$ is compact. Note that such groups are residually finite.
  The second construction described below allows us to find a
  decoupling family $(g_\eps)$ on $X$ where $g_\eps = \rho_\eps^2 g$
  is conformally equivalent to $g$. We then apply Main~Theorem~1 and
  obtain for each $n \in \N$ a metric $g_{\eps_n}$ such that the
  corresponding Laplacian has at least $n$ gaps.  In particular, the
  number of gaps is \emph{not} stable, even under uniform conformal
  changes of the metric. Note that the conformal factor $\rho_\eps$
  can be chosen in such a way that $\rho_\eps \to \rho_{\eps_0}$
  uniformly as $\eps \to \eps_0$ provided $\eps_0>0$.  Nevertheless,
  the band-gap structure remains invariant due to Main Theorem~3, once
  $\Gamma$ has a positive Kadison constant.
\end{consequence}

\begin{figure}
  \begin{center}
\begin{picture}(0,0)
  \includegraphics{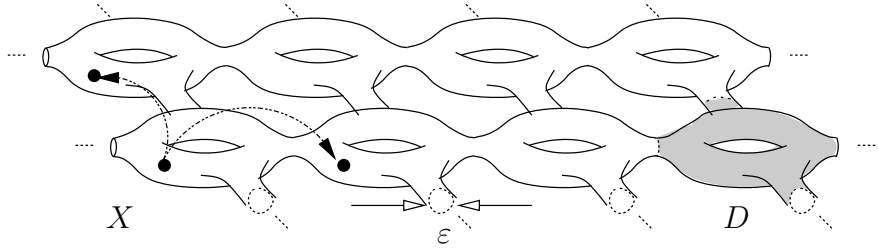}
\end{picture}%
\setlength{\unitlength}{4144sp}
\begin{picture}(5244,1501)(259,-695)
  \put(811,-556){$X$}
  \put(4501,-556){$D$}
  \put(2791,-646){$\eps$}
\end{picture}
    \caption{A covering manifold $X$ with fundamental domain $D$. The
      junctions between different translates of $D$ are of order
      $\eps$.}
    \label{fig:per-mfd}
  \end{center}
\end{figure}

\subsection{An outline of the argument}
\label{ssec:outline}
In the rest of the introduction we will present the main ideas of
the construction of the decoupling metrics and mention the strategy
for showing the existence of spectral gaps. 

The first construction starts from a compact Riemannian manifold $N$
of dimension $d \ge 2$ (for simplicity without boundary) and a group
$\Gamma$ with generators $\gamma_1, \dots, \gamma_r$. We choose $2r$
different points $x_1,y_1, \dots, x_r, y_r$. For each generator, we
endow $x_i$ and $y_i$ with a cylindrical end of radius and length of
order $\eps>0$ (by changing the metric appropriately on $D:=N
\setminus \{x_1, y_1, \dots, x_r, y_r\}$). If we join $\Gamma$ copies
of these decorated manifolds $(D,g_\eps)$ according to the Cayley
graph of $\Gamma$ associated to $\gamma_1, \dots, \gamma_r$, we obtain a
$\Gamma$-covering $X \to M$ with a decoupling family of metrics
$(g_\eps)$ (cf.~Figure~\ref{fig:per-mfd}).

The second construction starts with an arbitrary covering $(X,g) \to
(M,g)$ (with compact quotient) of dimension $d\ge 3$ and changes the
metric conformally, i.e.\ $g_\eps := \rho_\eps^2 g$, in such a way,
that $\rho_\eps$ is still periodic and of order $\eps$ close to the
boundary of a fundamental domain $D$; more details can be found in
Section~\ref{sec:construye}.  In the case of abelian coverings these
constructions have already been used in~\cite{post:03a}.

Once the construction of the family of decoupling metrics $(g_\eps)$
has been done, the strategy to show the existence of spectral gaps
goes as follows.  We consider first the Dirichlet $(+)$ and Neumann
$(-)$ eigenvalues $\EWDN k (\eps)$ of the Laplacian on the fundamental
domain $(D,g_\eps)$.  One can show that $\EWDN k (\eps)$ converges to
the eigenvalues $\EW k(0)$ of the Laplacian on the limit manifold
$(N,g)$ (see \cite{post:03a} and references therein).  In other words,
the Dirichlet-Neumann intervals
\begin{equation*}
  I_k(\eps) := [\EWN k (\eps), \EWD k (\eps)]
\end{equation*}
converge to a point as $\eps\to 0$. Therefore, if $\eps$ is small
enough, the union
\begin{equation*}
  I(\eps) := \bigcup_{k \in \N} I_k(\eps)
\end{equation*}
is a closed set having at least $n$ gaps, i.e.\ $n+1$ components as a
subset of $[0, \infty)$.

The rest of the argument depends on the properties of the covering
group $\Gamma$:

\begin{enumerate}
\item For abelian groups $\Gamma_{\mathrm{ab}}$, the inclusion $\spec
  \laplacian {(X, g_\eps)} \subset I(\eps)$ is given by the Floquet
  theory (cf.~Section~\ref{sec:floquet} or \cite{kuchment:93,
    sunada:88}).  Basically, one shows that $\laplacian {(X,g_\eps)}$
  is unitary equivalent to a direct integral of operators on
  $(D,g_\eps)$ acting on $\rho$-equivariant functions, where $\rho$
  runs through the set of irreducible unitary representations
  $\widehat \Gamma_{\mathrm{ab}}$ (characters).  Note that in the
  abelian case all $\rho$ are one-dimensional and $\widehat
  \Gamma_{\mathrm{ab}}$ is homeomorphic to (disjoint copies of) the
  torus $\Torus^r$.  The Min-max principle ensures that the $k$-th
  eigenvalue of the equivariant operator lies in $I_k(\eps)$.
  
\item If the group is non-abelian but still has only
  finite-dimensional irreducible representations, then one can show
  that the spectrum of the $\rho$-equivariant Laplacian is still
  included in $I(\eps)$.  In this case the (non-abelian) Floquet
  theory guarantees again that $\spec\laplacian {(X, g_\eps)} \subset
  I(\eps)$. The class of groups which satisfy the previous condition
  are type~I groups, i.e finite extensions of abelian groups. These
  groups have a dual object $\widehat \Gamma$ which is a nice measure
  space (\emph{smooth} in the terminology
  of~\cite[Chapter~2]{mackey:76}).
  
\item If the group is \emph{residually finite} (a much wider class of
  groups including type I groups), then one can construct a so-called
  \emph{tower of coverings} consisting of finite coverings $M_i \to M$
  ``converging'' to the original covering $X \to M$. The inclusion of
  the spectrum of $\laplacian {(X,g_\eps)}$ in the closure of the
  union over all spectra of $\laplacian {(M_i,g_\eps)}$ was shown
  in~\cite{ass:94,adachi:95}.  For the \emph{finite} coverings $M_i
  \to M$ we again have the inclusion $\spec {\laplacian{(M_i,g_\eps)}}
  \subset I(\eps)$.
\item For non-amenable groups (i.e.\ groups, for which $\spec
  \laplacian {(M,g_\eps)}$ is not included in $\spec \laplacian
  {(X,g_\eps)})$, cf.~Remark~\ref{rem:amenable}, we have to assure
  that any of the intervals $I_k(\eps)$ intersects $\spec \laplacian
  X$ non-trivially. This will be done in~Theorem~\ref{thm:spectrum}.
\end{enumerate}

\subsection*{Organisation of the paper}
In the following section we set up the problem, present the
geometrical context and state some results and conventions that will
be needed later. In Section~\ref{sec:construye} we present in detail
the two procedures for constructing covering manifolds with a
decoupling family of metrics.  In this case the set $I(\eps)$ defined
above will have at least a prescribed finite number of spectral gaps.
Each procedure is well adapted to a given initial geometrical context
(cf.~Remark~\ref{ExplainMethods} as well as
Examples~\ref{ex:fund.group} and \ref{ex:heisenberg}).  In
Section~\ref{sec:floquet} we show the inclusion of the spectrum of
equivariant Laplacians into the union of the Dirichlet-Neumann
intervals $I_k(\eps)$ and review briefly the Floquet theory for
non-abelian groups.  The Floquet theory is applied in
Section~\ref{sec:type.I} for coverings with type~I groups.  In
Section~\ref{sec:res.fin} we study a class of covering manifolds with
residually finite groups.  In Section~\ref{sec:kadison} we consider
residually finite groups $\Gamma$ that in addition have a positive
Kadison constant.  In Section~\ref{sec:examples} we illustrate the
results obtained with some classes of examples and point out their
mutual relations.  Subsection~\ref{sec:OpenQuestion} contains an
interesting example of a covering with an amenable, \emph{not}
residually finite group which cannot be treated with our methods.  We
expect though that in this case one can still generate spectral gaps
by the construction presented in Section~\ref{sec:construye}. Finally,
we conclude mentioning several possible applications for our results.

%
\section{Geometrical preliminaries: covering manifolds and Laplacians}
%
\label{sec:prelim}

We begin fixing our geometrical context and recalling some results
that will be useful later on. We denote by $X$ a \emph{non-compact}
Riemannian manifold of dimension $d \ge 2$ with a metric $g$.  We also
assume the existence of a finitely generated (infinite) discrete group
$\Gamma$ of isometries acting \emph{properly discontinuously} and
\emph{cocompactly} on $X$, i.e.\ for each $x \in X$ there is a
neighbourhood $U$ of $x$ such that the sets $\gamma U$ and $\gamma'U$ are
disjoint if $\gamma \ne \gamma'$ and $M:=X/\Gamma$ is compact. Moreover, the
quotient $M$ is a Riemannian manifold which also has dimension $d$ and
is locally isometric to $X$.  In other words, $\map \pi X M$ is a
\emph{Riemannian covering space} with covering group $\Gamma$.  We
call such a manifold \emph{$\Gamma$-periodic} or simply \emph{periodic}.
All groups $\Gamma$ appearing in this paper will satisfy the preceding
properties.

We also fix a \emph{fundamental domain} $D$, i.e.\ an open set $D
\subset X$ such that $\gamma D$ and $\gamma' D$ are disjoint for all $\gamma \ne
\gamma'$ and $\bigcup_{\gamma \in \Gamma} \gamma \overline D = X$. We always
assume that $\overline D$ is compact and that $\bd D$ is piecewise
smooth. If not otherwise stated we also assume that $D$ is connected.
Note that we can embed $D\subset X$ isometrically into the quotient
$M$.  In the sequel, we will not always distinguish between $D$ as a
subset of $X$ or $M$ since they are isometric.  For details we refer
to~\cite[\S6.5]{ratcliffe:94}.

As a prototype for an elliptic operator we consider the Laplacian
$\laplacian X$ on a Riemannian manifold $(X,g)$ acting on a dense
subspace of the Hilbert space $\Lsqr X$ with norm $\norm[X] \cdot$.
For the formulation of the Theorems~\ref{thm:gaps.type.I} and
\ref{thm:gaps.res.fin} and at other places, it is useful to denote
explicitly the dependence on the metric, since we deform the manifold
by changing the metric. In this case we will write $\laplacian
{(X,g)}$ for $\laplacian X$ or $\Lsqr {X,g}$ for $\Lsqr X$.

The positive self-adjoint operator 
$\laplacian X$ can be defined in terms of a
suitable qua\-dra\-tic form $q_X$ (see e.g.~\cite[Chapter~VI]{kato:95},
\cite{reed-simon-1} or \cite{davies:96}). Concretely we have
\begin{equation}
\label{def:quad.form}
  q_X(u):=\normsqr[X] {d u} = \int_X {|d u|^2},\quad
          u \in \Cci X
\end{equation}
where the integral is taken with respect to the volume density measure
of $(X,g)$.  In coordinates we write the pointwise norm of the
$1$-form $d u$ as
\begin{displaymath}
  |d u |^2(x)= \sum_{i,j} g^{ij}(x) \partial_i u(x) \, 
                                    \partial_j\overline {u(x)} ,
\end{displaymath}
where $(g^{ij})$ is the inverse of the metric tensor $(g_{ij})$ in a
chart. Taking the closure of the quadratic form we can extend $q_X$ onto
the Sobolev space
\begin{displaymath}
  \Sob X = \Sob {X,g} = \set{ u \in \Lsqr X}{ q_X(u) < \infty}.
\end{displaymath}
As usual the operator $\laplacian X$ is related with the quadratic form by the
formula $\iprod {\laplacian X u} u = q_X(u)$, $u \in \Cci X$. Since the metric
on $X$ is $\Gamma$-invariant, the Laplacian $\laplacian X$ (i.e.\ its
resolvent) commutes with the translation on $X$ given by
\begin{equation}
  \label{eq:transl}
   (T_\gamma u )(x) := u(\gamma^{-1}x), \quad u \in \Lsqr X, \gamma \in \Gamma.
\end{equation}
Operators with this property are called \emph{periodic}.

For an open, relatively compact subset $D \subset X$ with sufficiently smooth
boundary $\bd D$ (e.g.~Lipschitz) we define the Dirichlet (respectively,
Neumann) Laplacian $\laplacianD D$ (resp., $\laplacianN D$) via its quadratic
form $q_D^+$ (resp., $q_D^-$) associated to the closure of $q_D$ on $\Cci D$,
the space of smooth functions with compact support, (resp., $\Ci {\overline
  D}$, the space of smooth functions with continuous derivatives up to the
boundary).  We also use the notation $\Sobn D = \dom q_D^+$ (resp., $\Sob D =
\dom q_D^-$). Note that the usual boundary condition of the Neumann Laplacian
occurs only in the \emph{operator} domain via the Gau{\ss}-Green formula.
Since $\overline D$ is compact, $\laplacianD D$ has purely discrete spectrum
$\EWD k$, $k \in \N$.  It is written in ascending order and repeated according
to multiplicity.  The same is true for the Neumann Laplacian and we denote the
corresponding purely discrete spectrum by $\EWN k$, $k\in\N$.

One of the advantages of the quadratic form approach is that 
one can easily read off from the inclusion of domains an order relation 
for the eigenvalues. In fact, by the 
the \emph{min-max principle} we have
\begin{equation}
  \label{eq:min.max}
  \EWDN k =
  \inf_{L_k} \sup_{u \in L_k \setminus \{0\} }
      \frac {q_D^\pm(u)}{\normsqr u},
\end{equation}
where the infimum is taken over all $k$-dimensional subspaces $L_k$ of the
corresponding \emph{quadratic} form domain $\dom q_D^\pm$,
cf.~e.g.~\cite{davies:96}. Then the inclusion
\begin{equation}
  \label{eq:dom.mono}
   \dom q_D^+ = \Sobn D  \subset \Sob D = \dom q_D^- 
\end{equation}
implies the following important relation between the corresponding
eigenvalues
\begin{equation}
  \label{eq:ew.mono}
   \EWD k \ge \EWN k .
\end{equation}
This means, that the Dirichlet $k$-th eigenvalue is in general
larger than the $k$-th Neumann eigenvalue and this justifies 
the choice of the labels $+$, respectively, $-$.

%
\section{Construction of periodic manifolds}
%
\label{sec:construye}
In the present section we will give two different construction
procedures (labelled by the letters `A' and `B') for covering manifolds,
such that the corresponding Laplacian will have a prescribed finite
number of spectral gaps. In contrast with \cite{post:03a} (where only
abelian groups were considered) we will base the construction on the
specification of the quotient space $M=X/\Gamma$. By doing this, the
spectral convergence result in Theorem~\ref{thm:mfd.conv} becomes
manifestly independent of the fact whether $\Gamma$ is abelian or not.

Both constructions are done in two steps: first, we specify in two
ways the quotient $M$ together with a family of metrics $g_\eps$.
Second, we construct in either case the covering manifold with
covering group $\Gamma$ which has $r$ generators. In the last section
we will localise the spectrum of the covering Laplacian in certain
intervals given by an associated Dirichlet, respectively, Neumann
eigenvalue problem.  Some reasons for presenting two different
methods~(A) and~(B) are formulated in a final remark of this section.

\subsection{Construction of the quotient}
\label{ssec:quotient}

\begin{figure}[h]
 \begin{center}
\begin{picture}(0,0)
  \includegraphics{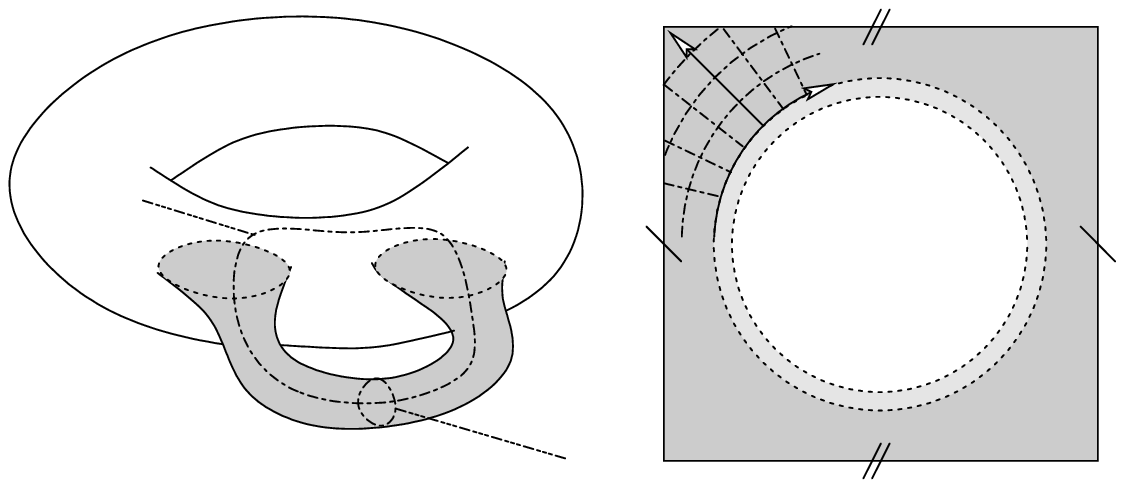}
\end{picture}
\setlength{\unitlength}{4144sp}
\begin{picture}(5335,2429)(361,-1861)
  \put(3601,-1800){(B)}
  \put(360,-1800){(A)}
  \put(3550,460){$r$}
  \put(4394,-44){$y$}
  \put(1391, 59){$R_\eps$}
  \put(360,-1550){$C=(0,1)\times\Sphere^{d-1}$}
  \put(360,250){$(N,g)$}
  \put(2000,-1650){$\beta_1$ (mid section)}
  \put(860,-400){$\alpha_1$}
  \put(4494,-854){$N$}
  \put(4970,178){$D \setminus N$}
  \put(4000,-1489){$\rho_\eps(x)=O(\eps)$}
  \put(4150,-1100){$\rho_\eps(x)=1$}
\end{picture}%
   \caption{Two constructions of a family of manifold $(M, g_\eps$), $\eps >
     0$: In both cases, the grey area has a length scale of order
     $\eps$ in all directions. (A)~We attach $r$ handles (here $r=1$)
     of diameter and length of order $\eps$ to the manifold $(N,g)$.
     We also denoted the two cycles $\alpha_1$ and $\beta_1$. (B)~We
     change the metric conformally to $g_\eps = \rho_\eps^2 g$. The
     grey area $D \setminus N$ (with Fermi coordinates in the upper
     left corner) shrinks conformally to a point as $\eps \to 0$
     whereas $N$ remains fixed. Note that the opposite sides of the
     square are identified (to obtain a torus as manifold $M$).}
   \label{fig:constr-mfd}
 \end{center}
\end{figure}

In the following two methods we define a family of Riemannian
manifolds $(M,g_\eps)$ that converge to a Riemannian manifold $(N,g)$
of the same dimension (cf.~Figure~\ref{fig:constr-mfd}). In each case
we will also specify a domain $D\subset M$ (in the following section
$D$ will become a fundamental domain of the corresponding covering):
\begin{enumerate}
\item[(1A)] \textbf{Attaching $r$ handles:} We construct the manifold
  $M$ by attaching $r$ handles diffeomorphic with $C := (0,1) \times
  \Sphere^{d-1}$ to a given $d$-dimensional compact orientable
  manifold $N$ with metric $g$.  For simplicity we assume that $N$ has
  no boundary.  Concretely, for each handle we remove two small discs
  of radius $\eps>0$ from $N$, denote the remaining set by $R_\eps$
  and identify $\{0\} \times \Sphere^{d-1}$ with the boundary of the
  first hole and $\{1\} \times \Sphere^{d-1}$ with the boundary of the
  second hole.  We denote by $D$ the open subset of $M$ where the mid
  section $\{1/2\} \times \Sphere^{d-1}$ of each handle is removed.
 
  One can finally define a family of metrics $(g_\eps)_\eps$,
  $\eps>0$, on $M$ such that the diameter and length of the handle is
  of order $\eps$ (see e.g.~\cite{post:03a,chavel-feldman:81}). In
  this situation the handles shrink to a point as $\eps\to 0$. Note
  that $(R_\eps, g)$ can be embedded isometrically into $(N,g)$, resp., 
  $(M, g_\eps)$. This fact will we useful for proving
  Theorem~\ref{thm:spectrum}.

\item[(1B)] \textbf{Conformal change of metric:} In the second
  construction, we start with an arbitrary compact $d$-dimensional
  Riemannian manifold $M$ with metric $g$. We consider only the case
  $d \ge 3$ (for a discussion of some two-dimensional examples
  see~\cite{post:03a}). Moreover, we assume that $N$ and $D$ are two
  open subsets of $M$ such that (i)~$\bd N$ is smooth, (ii)~$\overline
  N \subset D$, (iii)~$\overline D = M$ and (iv)~$D \setminus N$ can
  completely be described by Fermi coordinates (i.e.\ coordinates
  $(r,y)$, $r$ being the distance from $N$ and $y \in \bd N$) up to a
  set of measure $0$ (cf.~Figure~\ref{fig:constr-mfd}~(B)). The last
  assumption assures that $N$ is in some sense large in $D$.
  
  Suppose in addition, that $\map {\rho_\eps} M {(0,1]}$, $\eps>0$, is
  a family of smooth functions such that $\rho_\eps(x)=1$ if $x \in N$
  and $\rho_\eps(x)=\eps$ if $x \in M \setminus N$ and $\dist(x,\bd N)
  \ge \eps^d$. Then $\rho_\eps$ converges pointwise to the
  characteristic function of $N$. Furthermore, the Riemannian manifold
  $(M, g_\eps)$ with $g_\eps:= \rho_\eps^2 g$ converges to $(N,g)$ in
  the sense that $M \setminus N$ shrinks to a point in the metric
  $g_\eps$.
\end{enumerate}

Now we can formulate the following spectral convergence result which was
proven in~\cite{post:03a}:
\begin{theorem}
  \label{thm:mfd.conv}
  Suppose $(M,g_\eps)$ and $D \subset M$ are constructed as in
  parts~(1A) or~(1B) above. In Case~(1B) we assume in addition that $d
  \ge 3$. Then
  \begin{displaymath}
    \EWDN k (\eps) \to \EW k (0)
  \end{displaymath}
  as $\eps \to 0$ for each $k$. Here, $\EWDN k (\eps)$ denotes the
  $k$-th Dirichlet, resp., Neumann eigenvalue of the Laplacian on
  $(D,g_\eps)$ whereas $\EW k (0)$ is the $k$-th eigenvalue of $(N,g)$
  (with Neumann boundary conditions at $\bd N$ in Case~(1B)).
\end{theorem}

\subsection{Construction of the covering spaces}
\label{ssec:constr.cov.sp}

Given $(M,g_\eps)$ and $D$ as in the previous subsection, we will
associate a Riemannian covering $\map \pi {(X,g_\eps)} {(M, g_\eps)}$
with covering group $\Gamma$ such that $D$ is a fundamental
domain. Note that we identify $D \subset M$ with a component of the
lift $\widetilde D := \pi^{-1} (D)$. Moreover, $\Gamma$ is isomorphic
to a normal subgroup of the fundamental group $\pi_1(M)$.
\begin{enumerate}
\item[(2A)] Suppose that $\Gamma$ is a discrete group with $r$ generators
  $\gamma_1, \dots, \gamma_r$. We will construct a $\Gamma$-covering
  $(X,g_\eps) \to (M, g_\eps)$ with fundamental domain $D$ where $D$
  and $(M,g_\eps)$ are given as in Part~(1A) of the previous
  subsection. Roughly speaking, we glue together $\Gamma$ copies of
  $D$ along the handles according to the Cayley graph of $\Gamma$
  w.r.t.\ the generators $\gamma_1, \dots, \gamma_r$. For convenience
  of the reader, we specify the construction:
  
  The fundamental group of $M$ is given by $\pi_1(M) = \pi_1(N) *
  \Z^{*r}$ in the case $d \ge 3$. Here, $G_1*G_2$ denotes the free
  product of $G_1$ and $G_2$, and $\Z^{*r}$ is the free group in $r$
  generators $\alpha_1, \dots, \alpha_r$. If $d=2$ we know from the
  classification result for $2$-dimensional orientable manifolds that
  $N$ is diffeomorphic to an $s$-holed torus. In this case the
  fundamental group is given by
  \begin{equation}
    \label{eq:fund.group}
    \pi_1(M) = \langle \alpha_1, \beta_1, \dots,
  \alpha_{r+s}, \beta_{r+s} \mid [\alpha_1, \beta_1] \cdot \ldots \cdot
  [\alpha_{r+s},\beta_{r+s}] = e \rangle,
  \end{equation}
  where $[\alpha,\beta]:=\alpha \beta \alpha^{-1}\beta^{-1}$ is the usual
  commutator.  We may assume that $\alpha_i$
  represents the homotopy class of the cycle \emph{transversal} to the section
  of the $i$-th handle and that $\beta_i$ represents the section itself ($i=1,
  \dots, r$) (cf.~Figure~\ref{fig:constr-mfd}~(A)).
 
  One easily sees that there exists an epimorphism $\map \phi {\pi_1(M)} \Gamma$
  which maps $\alpha_i \in \pi_1(M)$ to $\gamma_i \in \Gamma$ ($i=1, \dots, r$) and
  all other generators to the unit element $e \in \Gamma$. 
  Note that this map is also well-defined in the case $d=2$,
  since the relation in~\eqref{eq:fund.group} is trivially satisfied in the
  case when the $\beta_i$'s are mapped to $e$.

  Finally, $\Gamma\cong \pi_1(M) / \ker \phi$, and 
  $X \to M$ is the associated covering with respect to the universal
  covering $\widetilde M \to M$ 
  (considered as a principal bundle with discrete fibre $\Gamma$) and
  the natural action of $\Gamma$ on $\pi_1(M)$.

  Then $X \to M$ is a normal $\Gamma$-covering with fundamental domain $D$
  constructed as in~(1A) of the preceding subsection. Here we use the
  fact that $\alpha_i$ is \emph{transversal} to the section of the
  handle in dimension~$2$.
  
\item[(2B)] Suppose $(X,g) \to (M,g)$ is a Riemannian covering 
  with fundamental
  domain $D$ such that $\bd D$ is piecewise smooth. Then $\overline D = M$,
  where we have embedded $D$ into the quotient,
  cf.~\cite[Theorem~6.5.8]{ratcliffe:94}. 
  According to~(1B) we can conformally change the metric on $M$, to produce
  a new covering $(X,g_\eps) \to (M,g_\eps)$ that satisfies the required 
  properties.

\end{enumerate}
In both cases, we lift for each $\eps >0$ the metric $g_\eps$ from $M$
to $X$ and obtain a Riemannian covering $(X,g_\eps) \to (M, g_\eps)$.
Note that the set $D$ specified in the first step of the previous
construction becomes a fundamental domain after the specification of
the covering in the second step.

The following statement is a direct consequence of the spectral convergence
result in Theorem~\ref{thm:mfd.conv}:
\begin{theorem}
  \label{thm:gaps}
  Suppose $(X,g_\eps) \to (M, g_\eps)$ ($\eps>0$) is a family of Riemannian
  coverings with fundamental domain $D$ constructed as in the 
  previous parts~(2A) or~(2B).
  Then for each $n \in \N$ there exists $\eps = \eps_n > 0$ such that
  \begin{equation}
    \label{eq:gaps}
    I(\eps):=\bigcup_{k \in \N} I_k(\eps), \qquad \text{with} \qquad
    I_k(\eps) := [\EWN k (\eps), \EWD k (\eps)],
  \end{equation}
  is a closed set having at least $n$ gaps, i.e.\ $n+1$ components as subset
  of $[0,\infty)$. Here, $\EWDN k (\eps)$ denotes the $k$-th Dirichlet, 
  resp., Neumann eigenvalue of the Laplacian on $(D,g_\eps)$.
\end{theorem}
\begin{proof}
  First, note that $\set{\EWDN k (\eps)}{ k \in \N}$, $\eps\geq 0$,
  has no finite accumulation point, since the spectrum is discrete.
  Second, Theorem~\ref{thm:mfd.conv} 
  shows that the intervals $I_k(\eps)$ reduce
  to the point $\{\EW k (0) \}$ as $\eps \to 0$.
  Therefore, $I(\eps)$ is a locally finite
  union of compact intervals, hence closed.
\end{proof}

\subsection{Existence of spectrum outside the gaps}
\label{ssec:ex.spec}

In the following subsection we will assure that each Neumann-Dirichlet
interval $I_k(\eps)$ contains at least one point of $\spec \laplacian
{(X,g_\eps)}$ provided $\eps$ is small enough. In our general setting
described below (cf.\ Theorems~\ref{thm:gaps.type.I}
and~\ref{thm:gaps.res.fin}) we will show the inclusion
\begin{equation}
  \label{eq:spec.incl}
  \spec \laplacian {(X,g_\eps)} \subset \bigcup_{k \in \N} I_k(\eps).
\end{equation}
It is a priori not clear that each $I_k(\eps)$ intersects the spectrum
of the Laplacian on $(X,g_\eps)$, i.e.\ that gaps in $\bigcup_{k \in
  \N} I_k(\eps)$ are also gaps in $\spec \laplacian {(X,g_\eps)}$. If
the covering group is amenable, the $k$-th eigenvalue of the Laplacian
on the quotient $(M,g_\eps)$ is always an element of $I_k(\eps) \cap
\spec (\laplacian X, g_\eps)$ (cf.~the argument in the proof of
Theorem~\ref{thm:gaps.type.I}). In general, this need not to be true.
Therefore, we need the following theorem which will be used in
Theorems~\ref{thm:gaps.res.fin} and~\ref{thm:band}:

\begin{theorem}
  \label{thm:spectrum}
  With the notation of the previous theorem, we have
  \begin{equation}
    \label{eq:spectrum}
    I_k(\eps) \cap \spec \laplacian {(X,g_\eps)} \ne \emptyset
  \end{equation}
  for all $k \in \N$.
\end{theorem}
We begin with a general criterion which will be useful to detect
points in the spectra of a parameter-dependent family of operators
using only its sesquilinear form. A similar result is also stated
in~\cite[Lemma~5.1]{krejcirik-kriz:05}.

Suppose that $H_\eps$ is a self-adjoint, non-negative, unbounded
operator in a Hilbert space $\HS_\eps$ for each $\eps>0$.  Denote by
$\HS_\eps^1 := \dom h_\eps$ the Hilbert space of the corresponding
quadratic form $h_\eps$ associated to $H_\eps$ with norm $\norm[1] u
:= (h_\eps(u) + \norm[\HS_\eps] u)^{1/2}$ and by $\HS_\eps^{-1}$ the
dual of $\HS_\eps^1$. Note that
$\map{H_\eps}{\HS_\eps^1}{\HS_\eps^{-1}}$ is continuous. In the next
lemma we characterise for each $\eps$ certain spectral points of
$H_\eps$.

\begin{lemma}
  \label{lem:char.spec}
  Suppose there exist a family $(u_\eps) \subset \HS_\eps^1$ and
  constants $\lambda \ge 0$, $c>0$ such that
  \begin{equation}
    \label{eq:char.spec}
    \norm[-1] {(H_\eps-\lambda)u_\eps} \to 0 \qquad \text{as} \qquad
    \eps \to 0
  \end{equation}
  and $\norm {u_\eps} \ge c > 0$ for all $\eps>0$, then there exists
  $\delta=\delta(\eps) \to 0$ as $\eps \to 0$ such that
  \begin{displaymath}
    \lambda + \delta(\eps) \in \spec H_\eps.
  \end{displaymath}
\end{lemma}
\begin{proof}
  Suppose that the conclusion is false. Then there exist a sequence
  $\eps_n \to 0$ and a constant $\delta_0>0$ such that
  \begin{displaymath}
    I_\lambda \cap \spec H_{\eps_n} = \emptyset \qquad \text{with} \qquad
    I_\lambda := (\lambda - \delta_0, \lambda + \delta_0)
  \end{displaymath}
  for all $n \in \N$.  Denote by $E_t$ the spectral resolution of
  $H_\eps$. Then
  \begin{multline*}
    \normsqr[-1] {(H_\eps - \lambda)u_\eps} =
    \Dint {\R_+ \setminus I_\lambda} {\frac{(t-\lambda)^2} {(t+1)}} 
                   {\iprod {E_tu_\eps} {u_\eps}} \\ \ge
    \frac {\delta_0^2} {\lambda + \delta_0 + 1}
    \Dint {\R_+ \setminus I_\lambda} 
          {} {\iprod {E_tu_\eps} {u_\eps}} \ge
    \frac {c \delta_0^2} {\lambda + \delta_0 + 1}
  \end{multline*}
  since $I_\lambda$ does not lie in the support of the spectral
  measure. But this inequality contradicts~\eqref{eq:char.spec}.
\end{proof}

\begin{remark}
  Eq.~\eqref{eq:char.spec} is equivalent to the inequality
  \begin{equation}
    \label{eq:char.spec2}
    |h_\eps(u_\eps, v_\eps) - \lambda \iprod {u_\eps} {v_\eps} | \le
    o(1) \norm[1] {v_\eps} \qquad \text{for all $v_\eps \in \HS_\eps^1$}
  \end{equation}
  as $\eps \to 0$. Note that $o(1)$ could depend on $u_\eps$.  The
  advantage of the criterion in the previous lemma is that one only
  needs to find a family $(u_\eps)$ in the domain of the quadratic
  form $h_\eps$.
\end{remark}

We will need the following lemma in order to define a cut-off function
with convergent $L_2$-integral of its derivative. Its proof is
straightforward.
\begin{lemma}
  \label{lem:cut-off}
  Denote by $h(r):= r^{-d+2}$ if $d \ge 3$ and $h(r) = \ln r$ if
  $d=2$. For $\eps \in (0,1)$ define
  \begin{equation}
  \label{eq:cut-off}
    \chi_\eps(r):=
    \begin{cases}
      0, & 0 < r \le \eps\\
      \frac{h(r)-h(\eps)}{h(\sqrt \eps) - h(\eps)}, & 
                        \eps \le r \le \sqrt \eps\\
      1, & \sqrt \eps \le r
    \end{cases}
  \end{equation}
  then $\chi_\eps \in \Sob{(0,1)}$ and
  \begin{displaymath}
    \normsqr {\chi_\eps'} := \Dint[1] 0 {|\chi_\eps'(r)|^2 r^{d-1}} r = o(1)
  \end{displaymath}
  as $\eps \to 0$.
\end{lemma}

Remember that $(N,g)$ is the unperturbed manifold as in
Figure~\ref{fig:constr-mfd}. In Case~A of
Subsection~\ref{ssec:quotient}, we denoted by $R_\eps$ the manifold
$N$ with a closed ball of radius $\eps$ removed around each point
where the handles have been attached (note that $R_\eps$ is also
contained in $D$) and denote by $(r,y)$ the polar coordinates around
such a point ($r=\eps$ corresponds to a component of $\bd R_\eps$).

\begin{proof}[Proof of Theorem~\ref{thm:spectrum}]
  Let $\phi$ be the $k$-th eigenfunction of the limit operator
  $\laplacian N$ with eigenvalue $\lambda=\lambda_k(0)$. We will treat
  Cases~A and~B of Subsection~\ref{ssec:quotient} separately.

  \noindent
  (3A) Set $u_\eps(r,y) := \chi_\eps(r) \phi(r,y)$ in the polar
  coordinates described above and $u_\eps := \phi$ on $R_{\sqrt
    \eps}$. Now, $\normsqr[R_{\sqrt \eps}] {\phi} \ge c$ since
  $\normsqr[R_{\sqrt \eps}] \phi \to \normsqr[N] \phi>0$ as $\eps \to
  0$. In addition, $u_\eps \in \Sobn {R_\eps} \subset \Sob {X,g_\eps}$
  and
    \begin{multline*}
      |\iprod {du_\eps} {dv_\eps} - \lambda \iprod {u_\eps} {v_\eps}| \\ =
      \Bigl| \int_{R_\eps} \bigl[\iprod {d\phi} {d(\chi_\eps v_\eps)}
                 -\lambda \phi \overline{\chi_\eps v} \bigr] +
        \int_{R_\eps} \phi \iprod {d\chi_\eps}{dv_\eps} -
        \int_{R_\eps} \overline v \iprod {d\phi}{d\chi_\eps}
      \Bigr|
    \end{multline*}
    for all $v_\eps \in \Sob {D_\eps}$.  Now the first integral
    vanishes since $\phi$ is the eigenfunction with eigenvalue
    $\lambda$ on $N$.  Note that $\chi_\eps v \in \Sobn {R_\eps}$ can
    be interpreted as function in $\Sob N$. The second and third
    integral can be estimated from above by
    \begin{displaymath}
     \sup_{x \in N} \bigl[|\phi(x)| + |d\phi(x)| \bigr]
       \norm {\chi_\eps'} \norm[1] {v_\eps} = o(1) \norm[1] {v_\eps}
    \end{displaymath}
    since $\phi$ is a smooth function on an $\eps$-independent space
    and due to Lemma~\ref{lem:cut-off}.
  
   \noindent
   (3B) Set $u_\eps := \phi$ on $N$ and $u_\eps(r,y):=\widetilde
   \chi_\eps(r) \phi(0,y)$, $r>0$, i.e.\ on $D \setminus N$ with
   $\widetilde \chi_\eps(r):=\chi_\eps(\sqrt \eps + \eps^d - r)$,
   where $\chi_\eps$ is defined in~\eqref{eq:cut-off} with $d=2$. Note
   that $\widetilde \chi_\eps'(r) \ne 0$ only for those $r=\dist(x,\bd
   N)$ where the conformal factor $\rho_\eps(x)=\eps$. Now, $u_\eps
   \in \Sobn {D,g_\eps} \subset \Sob {X,g_\eps}$. Furthermore, for
   $v_\eps \in \Sob {D,g_\eps}$ we have
    \begin{multline*}
      |\iprod {du_\eps} {dv_\eps} - \lambda \iprod {u_\eps} {v_\eps}| \le
      \int_{D \setminus N}
      \Bigl[ 
         \bigl|
             \widetilde \chi_\eps'(r)\phi(0,y) \partial_r v_\eps
         \bigr| \rho_\eps^{d-2} \\ +
         \bigl|
             \widetilde \chi_\eps (r) \iprod {d_y\phi(0,y)} {d_y v_\eps}
         \bigr| \rho_\eps^{d-2} +
         \lambda \widetilde \chi_\eps(r) |\phi(0,y) v_\eps| \rho_\eps^d
      \Bigr] \, \mathrm dr \, \mathrm dy \\ \le
       C
       \Bigl[
       \Bigl( 
         \int\limits_{\eps^d}^{\sqrt \eps + \eps^d - \eps} 
               |\widetilde \chi_\eps'(r)|^2 \eps^{d-2} \,\mathrm dr 
       \Bigr)^{\frac 12} \\+
       \Bigl( 
         \int\limits_0^{\sqrt \eps} 
               |\widetilde \chi_\eps(r)|^2 \rho_\eps^{d-2} \, \mathrm dr 
       \Bigr)^{\frac 12} +
       \Bigl( 
         \int\limits_0^{\sqrt \eps} 
               |\widetilde \chi_\eps(r)|^2 \rho_\eps^d \, \mathrm dr 
       \Bigr)^{\frac 12} 
       \Bigr] \norm[1]{v_\eps}
     \end{multline*}
     where we have used that $\phi$ is the Neumann eigenfunction on
     $N$. Furthermore, $C$ depends on the supremum of $\phi$ and
     $d\phi$ and on $\lambda$. Note that the conformal factor
     $\rho_\eps$ equals $\eps$ on the support of $\widetilde
     \chi_\eps'$, therefore, the first integral converges to $0$ since
     $d \ge 3$. Finally, estimating $\widetilde \chi_\eps$ and
     $\rho_\eps$ by $1$, the second and third integral are bounded by
     $\eps^{1/4}$.
\end{proof}

We finally can define formally the meaning of ``decoupling'':
\begin{definition}
  We call a family of metrics $(g_\eps)_\eps$ on $X \to M$
  \emph{decoupling}, if the conclusions of Theorems~\ref{thm:gaps}
  and~\ref{thm:spectrum} hold, i.e., if there exists a fundamental
  domain $D$ such that for each $n$ there exists $\eps_n>0$ such that
  $I(\eps_n)$ in \eqref{eq:gaps} has at least $n+1$ components and
  if~\eqref{eq:spectrum} holds for all $k \in \N$.
\end{definition}

\begin{remark}
\label{ExplainMethods}
In the present section we have specified two constructions of
decoupling families of metrics on covering manifolds, such that the
corresponding Laplacians will have at least a prescribed number of
spectral gaps (cf.~Sections~\ref{sec:type.I} and~\ref{sec:res.fin}).
The construction specified in method~(A) is feasible for every given
covering group $\Gamma$ with $r$ generators. Note that this method
produces fundamental domains that have smooth boundaries (see
e.g.~Example~\ref{ex:fund.group} below).

  The construction in~(B) applies for every given Riemannian
  covering $(X,g) \to (M,g)$, since, by the procedure described,
  one can modify conformally this covering in order 
  to satisfy the spectral convergence
  result of Theorem~\ref{thm:mfd.conv}
  (cf.~Example~\ref{ex:heisenberg}).
\end{remark}

%
\section{Floquet theory for non-abelian groups}
%
\label{sec:floquet}

The aim of the present section is to state a spectral inclusion result
(cf.~Theorem~\ref{thm:spec.incl}) and the direct integral
decomposition of $\laplacian X$ (cf.~Theorem~\ref{thm:floquet}) for
certain \emph{non-abelian} discrete groups $\Gamma$.  These results will
be used to prove the existence of spectral gaps in the situations
analysed in the next two sections.  A more detailed presentation of
the results in this section may be found in \cite{lledo-post:07}.

\subsection{Equivariant Laplacians}
\label{ssec:equiv.lapl}
We will introduce next a new operator that lies ``between'' the
Dirichlet and Neumann Laplacians and that will play an important role
in the following.  Suppose $\rho$ is a unitary representation of the
discrete group $\Gamma$ on the Hilbert space $\HS$, i.e.\ $\map \rho
\Gamma{\Unitary \HS}$ is a homomorphism. We fix a fundamental domain
$D$ for the $\Gamma$-covering $X \to M$.

We now introduce the space of smooth $\rho$-equivariant functions
\begin{equation}
  \label{def:equiv.fct}
  \CiR {D,\HS} := \set {h \restr D}
    {h \in \Ci {X, \HS}, \quad h(\gamma x) = \rho_\gamma h(x), \quad  
         \gamma \in \Gamma, x \in X}.
\end{equation}  
This definition coincides with the usual one for abelian groups,
cf.~\cite{lledo-post:07}. Note that we need \emph{vector-valued}
functions $\map h X \HS$ since the representation $\rho$ acts on the
Hilbert space $\HS$, which, in general, has dimension greater than
$1$.

We define next the so-called \emph{equivariant Laplacian} (w.r.t.\ the
representation $\rho$) on $\Lsqr {D,\HS} \cong \Lsqr D \otimes \HS$:
Let a quadratic form be defined by
\begin{equation}
\label{eq:quad.form}
  \normsqr[D] {dh} := \Dint D {\normsqr[\HS] {dh(x)}} {X(x)}
\end{equation}
for $h \in \CiR {D,\HS}$, where the integrand is locally given by
\begin{displaymath}
    \normsqr[\HS] {dh(x)} = 
    \sum_{i,j} g^{ij}(x) \, \iprod[\HS] {\partial_i h(x)} {\partial_j h(x)},
     \qquad x \in D.
\end{displaymath}
This generalises Eq.~\eqref{def:quad.form} to the case of vector-valued
functions. We denote the domain of the closure of the quadratic form by
$\SobR{D,\HS}$. The corresponding non-negative, 
self-adjoint operator on $\Lsqr {D,\HS}$, the
\emph{$\rho$-equivariant Laplacian}, will be denoted by $\laplacianR {D,\HS}$
(cf.~\cite[Chapter~VI]{kato:95}).

\subsection{Dirichlet-Neumann bracketing}
\label{ssec:dir-neu}
We study in this section the spectrum of a $\rho$-equivariant
Laplacian $\Delta^\rho$ associated with a finite-dimensional
representation $\rho$.  In particular, we show that $\spec
\Delta^\rho$ is contained in a suitable set determined by the spectrum
of the Dirichlet and Neumann Laplacians on $D$.  The key ingredient in
dealing with non-abelian groups is the observation that this set is
\emph{independent} of $\rho$.

We begin with the definition of certain operators acting in $\Lsqr
{D,\HS}$ and its eigenvalues. We denote by $\EWN m (\HS)$, $\EWR m
(\HS)$, resp., $\EWD m (\HS)$ the $m$-th eigenvalue of the operator
$\laplacianN {D,\HS}$, $\laplacianR {D,\HS}$, resp., $\laplacianD {D,
  \HS}$ corresponding to the quadratic form~\eqref{eq:quad.form} on
$\Sobn {D,\HS}$, $\SobR {D,\HS}$, resp., $\Sob {D,\HS}$.  Recall that
$\Sobn {D,\HS}$ is the $\Sobsymb^1$-closure of the space of smooth
functions $\map h D \HS$ with support away from $\bd D$ and $\Sob
{D,\HS}$ is the closure of the space of smooth functions with
derivatives continuous up to the boundary.

The proof of the next lemma follows, as in the abelian case
(cf.~Eqs.~\eqref{eq:dom.mono} and~\eqref{eq:ew.mono}), from the
reverse inclusions of the quadratic form domains
\begin{equation}
  \label{eq:dom.mono.2}
  \Sob {D,\HS} \supset \SobR {D,\HS} \supset \Sobn {D,\HS}
\end{equation}
and the min-max principle~\eqref{eq:min.max}.

\begin{lemma}
  \label{lem:bracketing}
  We have
  \begin{displaymath}
   \EWN m (\HS)  \le \EWR m (\HS) \le \EWD m (\HS)
  \end{displaymath}
  for all $m \in \N$.
\end{lemma}

>From the definition of the quadratic form in the Dirichlet, resp., Neumann
case we have that the corresponding
vector-valued Laplacians are a direct sum of the scalar operators. Therefore
the eigenvalues of the corresponding vector-valued Laplace operators consist
of repeated eigenvalues of the scalar Laplacian. We can 
arrange the former in the following way:
\begin{lemma}
  \label{lem:dn.scalar}
  If $n:=\dim \HS < \infty$ then
  \begin{displaymath}
    \EWDN m (\HS) = \EWDN k, \qquad
    m=(k-1)n+1, \dots, kn,
  \end{displaymath}
  where $\EWDN k$ denotes the (scalar) $k$-th Dirichet/Neumann eigenvalue on
  $D$.
\end{lemma}
\begin{proof}
  Note that $\laplacianDN {D, \HS}$ is unitarily equivalent to an $n$-fold
  direct sum of the scalar operator $\laplacianDN D$ on $\Lsqr D$ since there
  is no coupling between the components on the boundary.
\end{proof}

Recall the definition of the intervals $I_k := [\EWN k, \EWD k]$ in
Eq.~\eqref{eq:gaps} (for simplicity, we omit in the following the
index $\eps$). From the preceding two lemmas we may collect the $n$
eigenvalues of $\laplacianR {D,\HS}$ which lie in $I_k$:
\begin{equation}
  \label{eq:band.rho}
  B_k(\rho) := \set {\EWR m (\HS)} 
                  {m=(k-1)n+1, \dots, kn}
          \subset I_k,  \qquad n := \dim \HS.
\end{equation}
Therefore, we obtain the following spectral inclusion for equivariant
Laplacians. This result will be applied in 
Theorems~\ref{thm:gaps.type.I} and \ref{thm:gaps.res.fin}
below.
\begin{theorem}
  \label{thm:spec.incl}
  If $\rho$ is a unitary representation on a finite-dimensional
  Hilbert space $\HS$ then
  \begin{displaymath}
    \spec \laplacianR {D, \HS} =
    \bigcup_{k \in \N} B_k(\rho) \subseteq 
    \bigcup_{k \in \N} I_k 
  \end{displaymath}
  where $\laplacianR {D, \HS}$ denotes the $\rho$-equivariant Laplacian.
\end{theorem}

\subsection{Non-abelian Floquet transformation}
\label{ssec:floquet}

Consider first the right, respectively, left regular representation $R$,
resp., $L$ on the Hilbert space $\lsqr \Gamma$:
\begin{equation}
  \label{def:reg.rep}
  (R_\gamma a)_\tg = a_{\tg \gamma}, \qquad
  (L_\gamma a)_\tg = a_{\gamma^{-1}\tg},  \quad\qquad 
     a = (a_\gamma)_\gamma \in \lsqr \Gamma, 
      \quad \gamma,\tg \in \Gamma.
\end{equation}
Using standard results we introduce the following unitary map (see
e.g., \cite[Section~3 and the appendix]{lledo-post:07} and
references cited therein)
\begin{equation}
  \label{eq:fourier}
  \map F {\lsqr \Gamma}{\OintZ {\HS(z)}}
\end{equation}
for a suitable measure space $(Z, \mathrm dz)$.  The map $F$ is a
generalisation of the Fourier transformation in the abelian case.
Moreover, it transforms the right regular representation $R$ into the
following direct integral representation
\begin{equation}
  \label{eq:reg.rep.trafo}
    \widehat R_\gamma = F R_\gamma F^* = \OintZ {R_\gamma(z)}, \qquad \gamma \in \Gamma.
\end{equation}

\begin{remark}
  \label{rem:meas.space}
  Let $\al R$ be the von Neumann algebra generated by all unitaries
  $R_\gamma$, $\gamma \in \Gamma$, i.e.
  \begin{equation}
    \label{eq:gen.vn.algebra}
    \al R = \set {R_\gamma} {\gamma \in \Gamma}'',
  \end{equation}
  where $\al R'$ denotes the commutant of $\al R$ in $\End {\lsqr
    \Gamma}$. Then we decompose $\al R$ with respect to a maximal abelian
  von Neumann subalgebra $\al A\subset \al R'$ (for a concrete example
  see Example~\ref{ex:dir.int}). The space $Z$ is the compact
  Hausdorff space associated, by Gelfand's isomorphism, to a
  \emph{separable} $C^*$-algebra $\al C$, which is strongly dense in
  $\al A$.  Furthermore, $\mathrm d z$ is a regular Borel measure on
  $Z$.  We may identify the algebra $\al A$ with $\Linfty {Z,\mathrm
    dz}$ and since it is maximal abelian, the fibre representations
  $R(z)$ are irreducible a.e.\ (see
  \cite[Section~14.8~ff.]{wallach:92}).
\end{remark}

The generalised Fourier transformation introduced in
Eq.~\eqref{eq:fourier} can be used to decompose $\Lsqr X$ into a
direct integral. In particular, we define for a.e.~$z \in Z$:
\begin{equation}
  \label{eq:floquet.short}
  (Uu)(z)(x) := \sum_{\gamma \in \Gamma} \,u(\gamma x)  R_{\gamma^{-1}}(z) v(z)  ,
\end{equation}
where $v:=F \delta_e \in \lsqr \Gamma$, $u \in \Cci X$ and $x \in D$.   
The map $U$ extends to a unitary map
\begin{displaymath}
  \map U {\Lsqr X} {\OintZ 
      {\Lsqr {D,\HS(z)}} \cong \OintZ {\HS(z)} \otimes \Lsqr D},
\end{displaymath}
the so-called \emph{Floquet} or \emph{partial Fourier transformation}.
Moreover, operators commuting with the translation $T$ on $\Lsqr X$
are decomposable, in particular, we can decompose $\laplacian X$ since
its resolvent commutes with all translations~\eqref{eq:transl}. 

We denote by $\CiEq {D, \HS(z)}$ the set of smooth $R(z)$-equivariant
functions defined in~\eqref{def:equiv.fct} and $\laplacianZ D$ is the
$R(z)$-equivariant Laplacian in $\Lsqr {D,\HS(z)}$.  One can show in
this context (cf.~\cite{sunada:88,lledo-post:07}):
\begin{theorem}
  \label{thm:floquet}
   The operator $U$ maps $\Cci X$ into
   $\OintZ {\CiEq {D,\HS(z)}}$. Moreover, $\laplacian X$ is unitary
   equivalent to $\OintZ {\laplacianZ D}$ and
  \begin{equation}
  \label{eq:spec.dir.int}
    \spec \laplacian X \subseteq
    \overline {\bigcup_{z \in Z} \spec \laplacianZ D}.
  \end{equation}
  If $\Gamma$ is amenable (cf.\ Remark~\ref{rem:amenable}), then we
  have equality in \eqref{eq:spec.dir.int}.
\end{theorem}

\begin{example}
  \label{ex:dir.int}
  Let us illustrate the above direct integral decomposition in the
  case of the free group $\Gamma = \Z * \Z$ generated by $\alpha$ and
  $\beta$. Let $A \cong \Z$ be the cyclic subgroup generated by
  $\alpha$. We can decompose the algebra $\mathcal R$ given
  in~\eqref{eq:gen.vn.algebra} w.r.t.\ the abelian algebra $\mathcal A
  := \set{L_a \in \mathcal L(\lsqr \Gamma)}{a \in A} \subset \mathcal
  R'$, and, in this case, we have $Z = \Sphere^1$. Since the set
  $\set{a \gamma a^{-1}}{a \in A}$ is infinite provided $\gamma \notin
  A$, the algebra is \emph{maximal} abelian in $\mathcal R'$ (i.e.
  $\mathcal A = \mathcal A' \cap \mathcal R'$), and therefore, each
  fibre representation $R(z)$ is irreducible in $\HS(z)$.
  Moreover, since $L_a \in \mathcal A'$ ($a \in A$) we can also
  decompose these operators w.r.t\ the previous direct integral.
  
  We can give a more concrete realisation of the abstract Fourier
  transformation $F=F_\Gamma$ (see e.g.~\cite[Section~19]{robert:83}):
  We interprete $\Gamma \to A \setminus \Gamma$ as covering space with
  abelian covering group $A$ acting on $\Gamma$ from the left; the
  corresponding translation action $T_a$ on $\lsqr \Gamma$ coincides
  with the left regular representation $L_a$ ($a \in A$). The
  (abelian) Floquet transformation $U=U_A$ gives a direct integral
  decomposition
  \begin{equation*}
    \map{F_\Gamma = U_A} 
           {\lsqr \Gamma} {\Oint {\widehat A} {\HS(\chi)} \chi},
  \end{equation*}
  where $\HS(\chi) \cong \lsqr {A \setminus \Gamma}$ is the space of
  $\chi$-equivariant sequences in $\lsqr \Gamma$. Note that $\HS(\chi)$
  is infinite dimensional. A straightforward calculation shows that
  \begin{equation*}
    R_\gamma \cong \Oint {\widehat A} {R_\gamma(\chi)} \chi
      \quad \text{and} \quad
    L_a      \cong \Oint {\widehat A} {L_a     (\chi)} \chi,
  \end{equation*}
  where $R_\gamma(\chi) u(\tg) = u(\tg \gamma)$ and $L_a(\chi) u(\tg)=
  \overline \chi(a) u(\tg)$ for $u \in \HS(\chi)$. Note that
  $L_\gamma$, $\gamma \notin A$, does not decompose into a direct
  integral over $Z$ since it mixes the fibres. Furthermore, one sees
  that $v = (U \delta_e)(\chi)$ is the \emph{unique} normalised
  eigenvector of $R_a(\chi)$ with eigenvalue $\chi(a)$. This follows
  from the fact that the set of cosets $\set{A\gamma a}{a \in A}
  \subset A\setminus \Gamma$ is infinite provided $\gamma \notin A$.
  From the previous facts one can directly check that each $R(\chi)$
  is an irreducible representation of $\Gamma$ in $\HS(\chi)$ and that
  these representations are mutually inequivalent. Finally, $R(\chi)$
  is also inequivalent to any irreducible component of the direct
  integral decomposition obtained from a different maximal abelian
  subgroup $B \ne A$.
\end{example}

%
\section{Spectral gaps for type~I groups}
%
\label{sec:type.I}

We will present in this section the first method to show that the
Laplacian of the manifolds constructed in Section~\ref{sec:construye}
with (in general \emph{non-abelian}) type~I covering groups have an
arbitrary finite number of spectral gaps.  We begin recalling the
definition of type~I groups in the context of discrete groups.

\begin{definition}
  \label{def:type.I}
  A discrete group $\Gamma$ is of \emph{type~I} if $\Gamma$ is a finite
  extension of an abelian group, i.e.\ if there is an exact sequence
   \begin{displaymath}
     0 \longrightarrow 
     A \longrightarrow
     \Gamma\longrightarrow
     \Gamma_0 \longrightarrow 
     0 ,
   \end{displaymath}
   where $A \lhd \Gamma$ is abelian and $\Gamma_0 \cong \Gamma/ A$ is a finite
   group.
\end{definition}

\begin{remark}
  \label{rem:type.I}
  \begin{enumerate}
  \item 
    \label{rem:type.I.i}
    In the previous definition we have used a simple characterisation
    of countable, \emph{discrete} groups of type~I due to Thoma,
    cf.~\cite{thoma:64}.  Moreover, all irreducible representations of
    a type~I group $\Gamma$ are finite-dimensional and have a uniform
    bound on the dimension (see~\cite{thoma:64,moore:72}). Therefore,
    the following properties are all equivalent: (a)~there is a
    uniform bound on the dimensions of irreducible representations of
    $\Gamma$, (b)~all irreducible representations of $\Gamma$ are
    finite-dimensional, (c) $\Gamma$ is a finite extension of an
    abelian group, (d)~$\Gamma$ is CCR (completely continuous
    representation, cf.~\cite[Ch.~14]{wallach:92}), (e)~$\Gamma$ is of
    type~I.  Recall also that $\Gamma$ is of type~I iff the von
    Neumann algebra $\al R$ generated by $\Gamma$
    (cf.~Eq.~\eqref{eq:gen.vn.algebra}) is of \emph{type~I}
    (cf.~\cite{kaniuth:69}).
  
    Note that for our application it would be enough if $\Gamma$ has a
    decomposition over a measure space $(Z,\mathrm d z)$ as in
    Remark~\ref{rem:meas.space} such that \emph{almost} every
    representation $\rho(z)$ is finite-dimensional. But such a group
    is already of type~I: indeed, if the set $\set {z \in Z} {\dim
      \HS(z) = \infty}$ has measure $0$, then it follows from
    \cite[Section~II.3.5]{dixmier:81} that the von Neumann Algebra
    $\mathcal R$ (cf.~Eq.~\eqref{eq:gen.vn.algebra}) is of type~I.  By
    the above equivalent characterisation this implies that $\Gamma$
    is of type~I.
  \item
    \label{rem:type.I.ii}
    The following criterion (cf.~\cite{kaniuth:69,kallman:70}) will
    be used in Examples~\ref{ex:heisenberg} and \ref{ex:free.group} to
    decide that a group is not of type~I: The von Neumann algebra $\al
    R$ is of type~II$_1$ iff $\Gamma_{\mathrm{fcc}}$ has infinite
    index in $\Gamma$.  Here,
    \begin{equation}
      \label{eq:fcc}
      \Gamma_{\mathrm{fcc}} :=
      \set{\gamma \in \Gamma} {C_\gamma \text{ is finite}}
    \end{equation}
    is the set of elements $\gamma\in\Gamma$ having finite conjugacy
    class $C_\gamma$.  In particular such a group is not of type~I.
    Even worse: Almost all representations in the direct integral
    decomposition~\eqref{eq:reg.rep.trafo} are of type~II$_1$
    (\cite[Section~II.3.5]{dixmier:81}) and therefore
    infinite-dimensional (see e.g.~Example~\ref{ex:dir.int}).
  \end{enumerate}
\end{remark}

\begin{remark}
  \label{rem:amenable}
  The notion of amenable discrete groups will be useful at different
  stages of our approach.  For a definition of \emph{amenability} of
  a discrete group $\Gamma$ see e.g.~\cite{day:57} or
  \cite{brooks:81}. We will only need the following equivalent
  characterisations: (a)~$\Gamma$ is amenable. (b)~$0 \in \spec
  \laplacian X$~\cite{brooks:81}.  (c)~$\spec \laplacian M \subset
  \spec \laplacian X$~\cite[Propositions~7--8]{sunada:88}. Here, $X
  \to M$ is a covering with covering group $\Gamma$. Note that
  discrete type~I groups are amenable since they are finite extensions
  of abelian groups (extensions of amenable groups are again amenable,
  cf.~\cite[Section~4]{day:57}).
  
  We want to stress that Theorem~\ref{thm:spectrum} is no
  contradiction to the fact that $\Gamma$ is amenable iff $0 \in \spec
  \laplacian {(X,g_\eps)}$ although the first interval
  $I_1(g_\eps)=[0,\EWD k(g_\eps)]$ tends to $0$ as $\eps \to 0$.  Note
  that we have only shown that $I_1(g_\eps) \cap \spec \laplacian
  {(X,g_\eps)} \ne \emptyset$ and \emph{not} $0=\EW 1(M,g_\eps) \in
  \spec \laplacian {(X,g_\eps)}$ which is only true in the amenable
  case.
\end{remark}

The \emph{dual of $\Gamma$}, which we denote by $\hG$, is the set of
equivalence classes of unitary irreducible representations of $\Gamma$. We
denote by $[\rho]$ the (unitary) equivalence class of a unitary
representation $\rho$ on $\HS$.  Note that the spectrum of a
$\rho$-equivariant Laplacian and $\dim \HS$ only depend on the
\emph{equivalence class} of $\rho$.

If $\Gamma$ is of type~I, then the dual $\hG$ becomes a nice measure space
(``smooth'' in the terminology of \cite[Chapter~2]{mackey:76}).
Furthermore, we can use $\hG$ as measure space in the direct integral
decomposition defined in Subsection~\ref{ssec:floquet}. In particular,
combining the results of Section~\ref{sec:prelim}
and~\ref{sec:floquet} we obtain the main result for type~I
groups:
\begin{theorem}
  \label{thm:gaps.type.I}
  Suppose $X \to M$ is a Riemannian $\Gamma$-covering with fundamental
  domain $D$, where $\Gamma$ is a type~I group and denote by $g$ the
  Riemannian metric on $X$. Then
  \begin{displaymath}
    \spec \laplacian {(X,g)} \subset \bigcup_{k \in \N} I_k(g), 
    \qquad\mathrm{and}\qquad
    I_k(g) \cap \spec \laplacian{(X,g)} \ne \emptyset, \quad k \in \N,
  \end{displaymath}
  where $I_k(g):=[\EWN k (D,g), \EWD k (D,g)]$ is the
  Neumann-Dirichlet interval defined as in~\eqref{eq:gaps}.  In
  particular, for each $n \in \N$ there exists a metric $g=g_{\eps_n}$
  constructed as in Subsection~\ref{ssec:constr.cov.sp} such that
  $\spec \laplacian {(X,g)}$ has at least $n$ gaps, i.e.\ $n+1$
  components as subset of $[0, \infty)$.
\end{theorem}
\begin{proof}
  We have
  \begin{displaymath}
    \spec \laplacian X =
    \overline {\bigcup_{[\rho] \in \hG} \spec \laplacianR {D, \HS}}
    \subseteq  \overline {\bigcup_{k \in \N} I_k(g)}
    = \bigcup_{k \in \N} I_k(g),
  \end{displaymath}
  where we used the Theorem~\ref{thm:floquet} with $Z=\hG$ for the
  first equality and Theorem~\ref{thm:spec.incl} for the inclusion.
  Note that $\Gamma$ is amenable and that the latter theorem applies
  since all (equivalence classes of) irreducible representations of a
  type~I group are finite-dimensional
  (cf.~Remark~\ref{rem:type.I}~(\ref{rem:type.I.i})). The existence of
  gaps in $\bigcup_k I_k(g)$ follows from Theorem~\ref{thm:gaps}.
  
  Since $\Gamma$ is amenable, $\spec \laplacian M \subset \spec
  \laplacian X$ (cf.~(c) in Remark~\ref{rem:amenable}).  Moreover,
  from Eq.~\eqref{eq:band.rho} with $\rho$ the trivial representation
  on $\HS=\C$, we have that $\lambda_k(M) \in I_k$. Note that
  functions on $M$ correspond to functions on $D$ with periodic
  boundary conditions. Therefore, we have shown that every gap of the
  union $\bigcup_k I_k(g)$ is also a gap of $\spec \laplacian X$.
\end{proof}

%
\section{Spectral gaps for residually finite groups}
%
\label{sec:res.fin}

In this section, we present a new method to prove the existence of a
finite number of spectral gaps of $\laplacian X$.  The present
approach is applicable to so-called residually finite groups $\Gamma$,
which is a much larger class of groups containing type~I groups
(cf.~Section~\ref{sec:examples}). Roughly speaking, residually finite
means that $\Gamma$ has a lot of normal subgroups with finite index.
Geometrically, this implies that one can approximate the covering
$\map \pi X M$ with covering group $\Gamma$ by \emph{finite} coverings
$\map {p_i} {M_i} M$, where the $M_i$'s are compact.

Since the present section is central to the paper we will give for
completeness proofs of known results, namely for
Theorem~\ref{thm:res.fin} (see~\cite{ass:94,adachi:95}).

\subsection{Subcoverings and residually finite groups}
\label{ssec:sub.cov}
Suppose that $\map \pi X M$ is a covering with covering group $\Gamma$ (as
in Section~\ref{sec:prelim}). Corresponding to a normal subgroup $\Gamma_i
\lhd \Gamma$ we associate a covering $\map {\pi_i} X {M_i}$ such that
\begin{equation}
  \label{eq:sub.cov}
  \begin{diagram}
          &                             & X          &                   &\\
          &  \ldTo(2,2)^{\pi_i}_{\Gamma_i}  &        & \rdTo(2,2)^\pi_\Gamma& \\
     M_i  &                             & \rTo^{p_i}_{\Gamma/\Gamma_i} 
                                                             &         & M
  \end{diagram}
\end{equation}
is a commutative diagram. The groups under the arrows denote the
corresponding covering groups.

\begin{definition}
\label{def:res.fin}
  A (countable, infinite) discrete group $\Gamma$ is residually finite if
  there exists a monotonous decreasing sequence of normal subgroups
  $\Gamma_i \lhd \Gamma$ such that
\begin{equation}
  \label{eq:sub.groups}
  \Gamma=\Gamma_0 \rhd \Gamma_1 \rhd \dots \rhd \Gamma_i \rhd \cdots,  \quad
  \bigcap_{i \in \N} \Gamma_i = \{e\}  \quad \text{and} \quad
  \text{$\Gamma/\Gamma_i$ is finite.}
\end{equation}
Denote by $\mathfrak R \mathcal F$ the class of residually finite
groups.
\end{definition}
Suppose now that $\Gamma$ is residually finite. Then there exists a
corresponding sequence of coverings $\map {\pi_i} X {M_i}$ such that
$\map {p_i} {M_i} M$ is a \emph{finite} covering
(cf.~Diagram~\eqref{eq:sub.cov}). Such a sequence of covering maps is
also called \emph{tower of coverings}.

\begin{remark}
We recall also the following equivalent definitions of residually 
finite groups (see e.g.~\cite{magnus:69} or~\cite[Section~2.3]{robinson:82}).
\begin{enumerate}
\item A group $\Gamma$ is called \emph{residually finite} if for
  all $\gamma \in \Gamma\setminus \{e\}$ there is a group homomorphism
  $\map \Psi \Gamma G$ such that $\Psi(\gamma) \ne e$ and
  $\Psi(\Gamma)$ is a \emph{finite} group.
\item Let $\al F$ denote the class of finite groups. Then $\Gamma$ is
  residually finite, iff the so-called \emph{$\al F$-residual}
  \begin{equation}
    \label{eq:residual}
    \mathfrak{R}_{\al F}(\Gamma) := 
     \bigcap_{ \substack{N \lhd \Gamma\\\Gamma/N \in \al F}} N
\end{equation}  
is trivial, i.e.~$\mathfrak{R}_{\al F}(\Gamma)=\{e\}$. 
\end{enumerate}
\end{remark}

Next we give some examples for residually finite groups (cf.~the
survey article~\cite{magnus:69}):
\begin{example}
  \label{ex:res.fin}
  (i)~Abelian and finite groups are residually finite. (ii)~Free
  products of residually finite groups are residually finite, in
  particular, the free group in $r$ generators $\Z^{*r}$ is residually
  finite. (iii)~Finitely generated linear groups are residually finite
  (for a simple proof of this fact cf.~\cite{alperin:87}; a group is
  called \emph{linear} iff it is isomorphic to a subgroup of
  $\mathrm{GL}_n(\C)$ for some $n \in \N$.) In particular,
  $\mathrm{SL}_n(\Z)$, fundamental groups of closed, orientable
  surfaces of genus $g$ or, more generally, finitely generated
  subgroups of the isometry group on the hyperbolic space $\Hyp^d$ are
  residually finite.
\end{example}

Next we need to introduce a metric on the discrete space $\Gamma$:
\begin{definition}
  \label{def:word.met}
  Let $G$ be a set which generates $\Gamma$. The \emph{word metric}
  $d=d_G$ on $\Gamma$ is defined as follows: $d(\gamma,e)$ is the
  minimal number of elements in $G$ needed to express $\gamma$ as a word
  in the alphabet $G$; $d(e,e):=0$ and $d(\gamma,\tg) := d(\gamma
  \tg^{-1}, e)$.
\end{definition}

Geometrically, residually finiteness means that, given any compact set
$K \subset X$, there exists a finite covering $\map {p_i} {M_i} M$ and
a covering $\map {\pi_i} X {M_i}$ which is injective on $K$
(cf.~\cite{brooks:86}). This idea is used in the following lemma:
\begin{lemma}
  \label{lem:seq.fund.dom}
  Fix a fundamental domain $D$ for the covering $\map \pi X M$ and
  suppose that $\map {\pi_i} X {M_i}$ ($i \in \N$) is a tower of
  coverings as above. Then for each covering $\map {\pi_i} X {M_i}$
  there is a fundamental domain $D_i$ (not necessarily connected) such
  that
  \begin{displaymath}
    D_0 := D \subset D_1 \subset \dots \subset D_i \subset \cdots  
      \qquad \text{and} \qquad
    \bigcup_{i \in \N} D_i = X.
  \end{displaymath}
\end{lemma}
\begin{proof}
  It is enough to show the existence of a family of representants $R_i
  \subset \Gamma$ of $\Gamma/\Gamma_i$, $i\in\N$, satisfying
  \begin{displaymath}
    R_0 := \{e\} \subset R_1 \subset \dots \subset R_i \subset \cdots  
      \qquad \text{and} \qquad
    \bigcup_{i \in \N} R_i = \Gamma.
  \end{displaymath}
  In this case the fundamental domains are given explicitly by
  \begin{displaymath}
    D_i := \intr \bigcup_{r \in R_i} r^{-1} \overline D,
  \end{displaymath}
  where $\intr$ denotes the topological interior.
  
  Let $d$ be the word metric on $\Gamma$ with respect to the set of
  generators $G := \set{\gamma \in \Gamma}{\gamma\overline D \cap
    \overline D \ne \emptyset}$, which is naturally adapted to the
  fundamental domain $D$. Note that $G$ is finite and generates $\Gamma$
  since $\overline D$ is compact (cf.~\cite[Theorems~6.5.10
  and~6.5.11]{ratcliffe:94}).
  
  We choose a set of representants $R_i$ of $\Gamma/\Gamma_i$ that
  have minimal distance in the word metric to the neutral element,
  i.e.~if $r\in R_i$, then $d(r,e)\leq d(r\Gamma_i, e)$. Note that
  since $\Gamma_{i+1}\subset\Gamma_i$ we have $R_{i+1}\supset R_i$.
  To conclude the proof we have to show that every $\gamma \in \Gamma$ is
  contained in some $R_i$, $i \in \N$.  Since $\Gamma$ is finitely
  generated, there exists $n \in \N$ such that $\gamma\in B_{n} :=
  \{\gamma \in \Gamma\mid d(\gamma,e)\le n\}$.  Moreover, since $B_{2n}$ is
  finite and $\Gamma$ residually finite we also have
  $B_{2n}\cap\Gamma_i=\{e\}$ for $i$ large enough.  Therefore, any
  other element $\tg=\gamma \gamma_i^{-1}$ in the class $\gamma
  \Gamma_i$ with $\gamma_i \in \Gamma_i \setminus \{e\}$ has a distance
  greater than $n$, since
  \begin{displaymath}
    d(\tg, e) = d(\gamma\gamma_i^{-1},e) = d(\gamma, \gamma_i) \ge
    d(e,\gamma_i) - d(\gamma,e) > 2n - n = n.
  \end{displaymath}
  This implies that $\gamma\in R_i$ 
  by the minimality condition in the choice of the representants.
\end{proof}

\begin{theorem}
  \label{thm:res.fin}
  Suppose $\Gamma$ is residually finite with the associated sequence of coverings
  $\map {\pi_i} X {M_i}$ and $\map {p_i} {M_i} M$ as in~\eqref{eq:sub.cov}.
  Then
  \begin{displaymath}
    \spec \laplacian X \subseteq \overline {\bigcup_{i \in \N} \spec \laplacian
    {M_i}},
  \end{displaymath}
  and the Laplacian $\laplacian {M_i}$ w.r.t.\ the finite covering
  $\map {p_i} {M_i} M$ has discrete spectrum. Equality holds iff
  $\Gamma$ is amenable.
\end{theorem}
\begin{proof}
  (Cf.~\cite{adachi:95}) If $\lambda \in \spec \laplacian X$, then for each
  $\eps>0$ there exists $u \in \Cci X$ such that
  \begin{displaymath}
    \frac{\normsqr[X] {(\laplacian X - \lambda) u}} {\normsqr[X] u} < \eps.
  \end{displaymath}
  Applying Lemma~\ref{lem:seq.fund.dom} there is an $i=i(\eps)$ such that
  $\supp u \subset D_i$. Furthermore, since
  $D_i \hookrightarrow M_i=X/\Gamma_i$ is an
  isometry, $u$ can be written as the lift of a smooth $f$ on $M_i$,
  i.e.\ $f \circ \pi_i = u$. Therefore,
  \begin{displaymath}
    \frac{\normsqr[M_i] {(\laplacian {M_i} - \lambda) f}} {\normsqr[M_i] f} =
    \frac{\normsqr[X] {(\laplacian X - \lambda) u}} {\normsqr[X] u} < \eps,
  \end{displaymath}
  which implies $\lambda \in \overline {\bigcup_{i \in \N} \spec
    \laplacian {M_i}}$. Finally, since $M_i \to M$ is a finite
  covering and $M$ is compact, $\spec\laplacian {M_i}$ is discrete.
  For the second assertion cf.~\cite{adachi:95} or~\cite{ass:94}. One
  basically uses the characterisation due to \cite{brooks:81} that
  $\Gamma$ is amenable iff $0 \in \spec \laplacian X$ (cf.\ 
  Remark~\ref{rem:amenable}).
\end{proof}

Next we analyse the spectrum of the finite covering $M_i \to M$. Note that 
$D$ is also isometric 
to a fundamental domain for \emph{each} finite covering $M_i \to M$,
$i \in \N$.
\begin{lemma}
  \label{lem:fin.group}
  We have
  \begin{displaymath}
    \spec \laplacian {M_i} = 
       \bigcup_{[\rho] \in \widehat{G_i}}
                       \spec \laplacianR {D,\HS(\rho)},
  \end{displaymath}
  where $\Delta^\rho$ is the equivariant Laplacian introduced in
  Subsection~\ref{ssec:equiv.lapl} and $G_i := \Gamma/\Gamma_i$ is a finite
  group and $\widehat{G_i}$ its dual.
\end{lemma}
\begin{proof}
  Applying the results of Subsection~\ref{ssec:floquet} to the finite group
  $G_i$ and the finite measure space $Z:=\widehat{G_i}$ with the counting
  measure all direct integrals become direct sums. By Peter-Weyl's theorem
  (see e.g.~\cite[\S27.49]{hewitt-ross-2}) we also have
  \begin{displaymath}
    \map F {\lsqr {G_i}} 
       {\bigoplus_{[\rho] \in \widehat{G_i}} n(\rho) \HS(\rho)},
  \end{displaymath}
  where each multiplicity satisfies $n(\rho)=\dim \HS(\rho)<\infty$. 
  Finally,
  \begin{displaymath}
    \laplacian {M_i} \cong 
      \bigoplus_{[\rho] \in \widehat{G_i}} \laplacianR {D, \HS(\rho)}
  \end{displaymath}
  and the result follows.
\end{proof}

We now can formulate the main result of this section:
\begin{theorem}
  \label{thm:gaps.res.fin} 
  Suppose $X \to M$ is a Riemannian $\Gamma$-covering with fundamental
  domain $D$, where $\Gamma$ is a residually finite group and denote by
  $g$ the Riemannian metric on $X$.  Then
  \begin{displaymath}
    \spec \laplacian {(X,g)} \subset \bigcup_{k \in \N} I_k(g), \qquad
    I_k(g) \cap \spec \laplacian{(X,g)} \ne \emptyset, \quad k \in \N,
  \end{displaymath}
  where $I_k(g):=[\EWN k (D,g), \EWD k (D,g)]$ is defined as
  in~\eqref{eq:gaps}.  In particular, for each $n \in \N$ there exists
  a metric $g=g_{\eps_n}$, constructed as in
  Subsection~\ref{ssec:constr.cov.sp}, such that $\spec \laplacian
  {(X,g)}$ has at least $n$ gaps, i.e.\ $n+1$ components as subset of
  $[0, \infty)$.
\end{theorem}
\begin{proof}
  We have
  \begin{displaymath}
    \spec \laplacian X \subseteq 
    \overline {\bigcup_{i \in \N} \spec \laplacian {M_i}} =
    \overline {\bigcup_{\substack{i \in \N\\ [\rho] \in \widehat {G_i}}}
           \spec \laplacianR {D, \HS(\rho)}}
    \subseteq  
    \overline {\bigcup_{k \in \N} I_k(g) } = \bigcup_{k \in \N} I_k(g),
  \end{displaymath}
  where we used Theorem~\ref{thm:res.fin}, Lemma~\ref{lem:fin.group}
  and Theorem~\ref{thm:spec.incl}. Note that the latter theorem
  applies since all (equivalence classes of) irreducible
  representations of the finite groups $G_i$, $i\in\N$, are
  finite-dimensional. The existence of gaps in $\bigcup_k I_k(g)$
  follows from Theorem~\ref{thm:gaps}. Finally, by
  Theorem~\ref{thm:spectrum}, a gap of $\bigcup_k I_k(g)$ is in fact a
  gap of $\spec \laplacian X$.
\end{proof}

%
\section{Kadison constant and asymptotic behaviour}
%
\label{sec:kadison}

In the present section we will combine our main result stated in
Theorem~\ref{thm:gaps.res.fin} with some results by Sunada and
Br\"uning (cf.~\cite[Theorem~1]{sunada:92} or
\cite{bruening-sunada:92}), to give a more complete description of the
spectrum of the Laplacian $\laplacian X$, where $X \to M$ is the
$\Gamma$-covering constructed in Section~\ref{sec:construye}. For
this, we need a further definition:
\begin{definition}
  \label{def:kadison}
  Let $\Gamma$ be a finitely generated discrete group.
  The \emph{Kadison constant} of $\Gamma$ is defined as
  \begin{displaymath}
     C(\Gamma) := \inf \set{ \tr_\Gamma (P)} 
           {\text{$P$  non-trivial projection in 
                           $C^*_{\mathrm{red}}(\Gamma,\al K)$}},
  \end{displaymath}
  where $\tr_\Gamma (\cdot)$ is the canonical trace on
  $C^*_{\mathrm{red}}(\Gamma,\al K)$ , the tensor product of the
  reduced group $C^*$-algebra of $\Gamma$ and the algebra $\al K$ of
  compact operators on a separable Hilbert space of infinite dimension
  (see \cite[Section~1]{sunada:92} for more details.)
\end{definition}
In this section, we assume that $\Gamma$ is is residually finite and
has a strictly positive Kadison constant, i.e.~$C(\Gamma)>0$.  For
example, the free product $\Z^{*r} * \Gamma_1 * \dots * \Gamma_a$ with
finite groups $\Gamma_i$ satisfies both properties
(cf.~e.g.~\cite{magnus:69}, \cite[Appendix]{sunada:92}). Another such
group is the fundamental group (cf.~Eq.~\eqref{eq:fund.group}) of a
(compact, orientable) surface of genus $g$
(see~\cite{marcolli-mathai:99}).

\begin{remark}
  Suppose that $K$ is an integral operator on $\Lsqr X$ commuting
  with the group action, having smooth kernel $k(x,y)$ and satisfying
  \begin{displaymath}
    k(x,y) = 0 \qquad \text{for all $x,y \in X$ with $d(x,y) \ge c$}
  \end{displaymath}
  for some constant $c>0$. Then $K$ can be interpreted as an element
  of $C_{\mathrm {red}}^*(\Gamma, \al K)$ and one can write the
  $\Gamma$-trace as
  \begin{displaymath}
    \tr_\Gamma K = \int_D k(x,x)\, dx
  \end{displaymath}
  (see \cite[Section~1]{sunada:92} as well as \cite{atiyah:76} for
  further details), where $D$ is a fundamental domain of $X \to M$.

  If we consider the spectral
  resolution of the Laplacian $\laplacian X \cong \Oint{} \lambda
  {E(\lambda)}$, then it follows that
  \begin{displaymath}
    E(\lambda_2) - E(\lambda_1) \in C_{\mathrm {red}}^*(\Gamma, \al K)
  \end{displaymath}
  if $\lambda_1 < \lambda_2$ and $\lambda_1, \lambda_2 \not\in \spec
  \laplacian X$ (cf.~\cite[Section~2]{sunada:92}).
\end{remark}

Denote by $\mathcal N(g,\lambda)$ the number of components of
$\spec\laplacian {(X,g)} \cap [0, \lambda]$.
From~\cite{bruening-sunada:92,sunada:92} we obtain the following
asymptotic estimate on $\mathcal N(g,\lambda)$:
\begin{theorem}
  \label{thm:lower.asym}
  Suppose $(X,g) \to (M,g)$ is a Riemannian $\Gamma$-covering where
  $\Gamma$ has a positive Kadison
  constant, i.e.\ $C(\Gamma)>0$ then
  \begin{equation}
  \limsup_{\lambda\to\infty} 
    \frac{\mathcal N(g, \lambda)} 
                   {(2\pi)^{-d} \omega_d \vol (M,g) \lambda^{d/2}}
    \le \frac 1 {C(\Gamma)}.
  \end{equation}
  In particular, the spectrum of $\laplacian X$ has band-structure,
  i.e.\ $\mathcal N(g,\lambda)<\infty$ for all $\lambda \ge 0$.
\end{theorem}
\begin{remark}
  \label{rem:bethe.sommer}
  Note that Theorem~\ref{thm:lower.asym} only gives an \emph{asymptotic}
  upper bound on the number of components of $\spec \laplacian X \cap
  [0,\lambda]$, not on the \emph{whole} spectrum itself.  Therefore,
  we have no assertion about the so-called \emph{Bethe-Sommerfeld
    conjecture} stating that the number of spectral gaps for a
  periodic operator in dimensions $d \ge 2$ remains \emph{finite}.
\end{remark}

Combining Theorem~\ref{thm:lower.asym} with our result on spectral gaps
we obtain more information on the spectrum and a \emph{lower}
asymptotic bound on the number of components:
\begin{theorem}
  \label{thm:band}
  Suppose $(X,g) \to (M,g)$ is a Riemannian $\Gamma$-covering where
  $\Gamma$ is a residually finite group and where $g=g_\eps$ is the
  family of decoupling metrics constructed in
  Section~\ref{sec:construye}. Then we have:
\begin{enumerate}
\item For each $n \in \N$ there exists $g=g_{\eps_n}$ such that $\spec
  \laplacian {(X,g)}$ has at least $n$ gaps. If in addition
  $C(\Gamma)>0$ then there exists $\lambda_0>0$ such that
  \begin{equation*}
    n +1 \le \mathcal N(g,\lambda) < \infty
  \end{equation*}
  for all $\lambda \ge \lambda_0$, i.e.\ $\spec {\laplacian{(X,g)}}$
  has band-structure.
\item Suppose in addition that the limit manifold $(N,g)$ has simple
  spectrum, i.e.\ all eigenvalues $\lambda_k(0)$ have multiplicity $1$
  (cf.~Theorem~\ref{thm:mfd.conv}). Then for each $\lambda \ge 0$
  there exists $\eps(\lambda)>0$ such that
  \begin{equation*}
    \liminf_{\lambda\to\infty} 
     \frac{\al N (g_{\eps(\lambda)}, \lambda)}
       {(2\pi)^{-d} \omega_d \vol (N,g) \lambda^{d/2}}
    \ge 1.
  \end{equation*}
  Here, $g_\eps$ denotes the metric constructed in
  Section~\ref{sec:construye}.
\end{enumerate}
\end{theorem}
\begin{proof}
  (i)~follows immediately from Theorems~\ref{thm:gaps.res.fin}
  and~\ref{thm:lower.asym}.  (ii)~Suppose $\lambda \notin \spec
  \laplacian N$, then $\lambda_k(0) < \lambda < \lambda_{k+1}(0)$ for
  some $k\in\N$.  Let $\eps=\eps(\lambda) \in (0,1]$ be the largest
  number such that $\al N(\lambda,g_\eps)$ is (at least) $k$, in other
  words, $\al N(\lambda, g_\eps) \ge k = \al N(\lambda, \laplacian N)$
  where the latter number denotes the number of eigenvalues of
  $\laplacian N$ below $\lambda$. We conclude with the Weyl theorem,
  \begin{displaymath}
  \lim_{\lambda\to\infty} 
     \frac{\al N (\lambda, \laplacian N)}
       {(2\pi)^{-d} \omega_d \vol (N,g) \lambda^{d/2}} = 1,
  \end{displaymath}
  where $\omega_d$ denotes the volume of the $d$-dimensional Euclidean
  unit ball.
\end{proof}

\sloppy To conclude the section we remark that generically,
$\laplacian {(N,g)}$ has simple spectrum (cf.~\cite{uhlenbeck:76}).
The assumption on the spectrum of $(N,g)$ is natural since $\mathcal
N(g,\lambda)$ counts the components without multiplicity.

%
\section{Examples}
%
\label{sec:examples}

%
%
\subsection{Relation between the approaches presented in 
  Sections~\ref{sec:type.I} and \ref{sec:res.fin}}
%
%
\label{Relation5.6} 

We begin comparing the two main approaches presented in this
paper which assure the existence of spectral gaps
(cf.~Sections~\ref{sec:type.I} and \ref{sec:res.fin}).

One easily sees from Definition~\ref{def:res.fin} that a \emph{finite}
extension of a residually finite group is again residually finite. In
particular, type~I groups are residually finite as finite extensions
of abelian groups (cf.\ Definition~\ref{def:type.I}).  Therefore, for
type~I groups one can also produce spectral gaps by the approximation
method with finite coverings introduced in Section~\ref{sec:res.fin}.
Nevertheless we believe that the direct integral method will be useful
when analysing further spectral properties:

\begin{example}
 \label{rem:ev.dep.cont}
 One of the advantages of the method described in
 Section~\ref{sec:type.I} is that one has more information about the
 bands.  Suppose $\Gamma$ is finitely generated and \emph{abelian},
 i.e.\ $\Gamma \cong \Z^r \oplus \Gamma_0$, where $\Gamma_0$ is the
 torsion subgroup of $\Gamma$. Then $\hG$ is the disjoint union of
 finitely many copies of $\Torus^r$.  From the continuity of the map
 $\rho \to \EWR k$ (cf.~\cite{bjr:99} or~\cite{sunada:90}), we can
 simplify the characterisation of the spectrum in
 Theorem~\ref{thm:floquet} and obtain
 \begin{equation}
   \label{eq:char.spec.ab}
   \spec \laplacian X = \bigcup_{k \in \N} B_k, \quad \text{where} \quad
   B_k := \set{\EWR k} {\rho \in \hG} \subseteq I_k,
 \end{equation}
 the $k$-th \emph{band}. Since $\hG$ is compact, $B_k$ is also
 compact, but in general, $B_k$ need not to be connected (recall that
 $\hG$ is connected iff $\Gamma$ is torsion free, i.e.\ $\Gamma= \Z^r$).
 Note also that $B_k$ has only finitely many components.  For
 non-abelian groups this approach may be generalised in the direction
 of Hilbert C*-modules (cf.~\cite{gruber:01}).
\end{example}

In principle one could also consider a combination of the methods of
Section~\ref{sec:type.I} and~\ref{sec:res.fin}: denote by $\al T_1$
the class of type~I groups and by $\mathfrak{R}\mathcal{T}_1$ the
class of \emph{residually type~I} groups, i.e.
$\Gamma\in\mathfrak{R}\mathcal{T}_1$ iff the $\al T_1$-residual
$\mathfrak{R}_{\mathcal{T}_1}(\Gamma)$ is trivial
(cf.~Eq.~\eqref{eq:residual}).  Similarly we denote by
$\mathfrak{R}\mathcal{F}$ the class of residually finite groups
(cf.~Definition~\ref{def:res.fin}).  If we consider a covering with a
group $\Gamma\in\mathfrak{R}\mathcal{T}_1$, then instead of the
\emph{finite} covering $\map{p_i} {M_i} M$ considered in
Eq.~\eqref{eq:sub.cov} we would have a covering with a type~I group.
For these groups, we can replace Lemma~\ref{lem:fin.group} by the
direct integral decomposition of Theorem~\ref{thm:floquet}.
Nevertheless the following lemma shows that the class of residually
finite and residually type~I groups coincide.

\begin{lemma}
  \label{lem:res.class}
  From the inclusion
  $\mathcal{F}\subset\mathcal{T}_1\subset\mathfrak{R}\mathcal{F}$ it
  follows that the corresponding residuals for the group $\Gamma$
  coincide, i.e.\ $\mathfrak R_{\al F}(\Gamma)=\mathfrak R_{\al T_1}
  (\Gamma)$.  Moreover,
  $\mathfrak{R}\mathcal{F}=\mathfrak{R}\mathcal{T}_1$.
\end{lemma}
\begin{proof}
  From the inclusion $\mathcal{F}\subset\mathcal{T}_1$ it follows
  immediately that $\mathfrak R_{\al F}(\Gamma)\supset \mathfrak
  R_{\al T_1} (\Gamma)$.  To show the reverse inclusion one uses the
  following characterisation: a group is residually $\al F$ iff it is
  a subcartesian product of finite groups
  (cf.~\cite[\S~2.3.3]{robinson:82}).  Finally, from the equality of
  the residuals it follows that
  $\mathfrak{R}\mathcal{F}=\mathfrak{R}\mathcal{T}_1$.
\end{proof}

%
%
\subsection{Examples with residually finite groups}
%
%
\label{ResFinGroups}

In the rest of this subsection we present several examples 
of residually finite groups which are not type~I. They
show different aspects of our analysis.

For the next example recall the construction~(A) described in
Section~\ref{sec:construye}.
\begin{example}[Fundamental groups of oriented, closed surfaces]
  \label{ex:fund.group}
  Suppose that $N:=\Sphere^2$ is the two-dimensional sphere with a
  metric such that $\laplacian N$ has simple spectrum
  (cf.~\cite{uhlenbeck:76} for the existence of such metrics).
  Suppose, in addition, that $M$ is obtained by adding $r$ handles to
  $N$ as described in Section~\ref{sec:construye}, Case~A.  The
  fundamental group $\Gamma$ of $M$ (cf.~Eq.~\eqref{eq:fund.group}
  with $s=0$) is residually finite (recall
  Example~\ref{ex:res.fin}~(iii)).  Moreover, from the proof of
  Proposition~2.16 in \cite{marcolli-mathai:99}, it follows that
  $\Gamma$ has a positive Kadison constant. Therefore,
  Theorem~\ref{thm:band} applies to the the universal cover $X
  :=\widetilde M \to M$ with the metric $g_\eps$ specified in
  Section~\ref{sec:construye}.
\end{example}

The following example uses the construction~(B) in
Section~\ref{sec:construye}.
\begin{example}[Heisenberg group]
  \label{ex:heisenberg}
  Let $\Gamma := H_3(\Z)$ be the \emph{discrete Heisenberg group},
  where $H_3(R)$ denotes the set of matrices
  \begin{equation}
    \label{eq:heisenberg}
    A_{x,y,z} := \begin{pmatrix}
      1 & x & y \\ 0 & 1 & z \\ 0 & 0 & 1
    \end{pmatrix}
  \end{equation}
  with coefficients $x,y,z$ in the ring $R$. A covering with group
  $\Gamma$ is given e.g.~by $X:=H_3(\R)$ with compact quotient
  $M:=H_3(\R)/H_3(\Z)$. Note that $X$ is diffeomorphic to $\R^3$.
  Clearly, $\Gamma$ is a finitely generated linear group and therefore
  residually finite (cf.\ Example~\ref{ex:res.fin}~(iii)). Now, by
  Theorem~\ref{thm:gaps.res.fin} one can deform conformally a
  $\Gamma$-invariant metric $g$ as in Case~(B) of
  Section~\ref{sec:construye}, such that $\spec \laplacian X$ has at
  least $n$ spectral gaps, $n\in\N$.

  In this case, $\Gamma$ is also amenable as an extension of amenable groups
  (cf.\ Remark~\ref{rem:amenable}). In fact,
  $\Gamma$ is isomorphic to the semi-direct product 
  $\Z\ltimes \Z^2$, where $1 \in \Z$ acts on $\Z^2$ by the matrix
  \begin{displaymath}
    \begin{pmatrix}
      1 & 1 \\ 0 & 1
    \end{pmatrix}.
  \end{displaymath}
  Therefore, we have equality in the
  characterisation of $\spec \laplacian X$ in
  Theorems~\ref{thm:floquet} and~\ref{thm:res.fin}.
  
  Note finally that the group $\Gamma$ is not of type~I since
  $\Gamma_{\mathrm {fcc}} = \set{A_{0,y,0}}{y \in \Z}$ has infinite
  index in $\Gamma$ (cf.\ 
  Remark~\ref{rem:type.I}~(\ref{rem:type.I.ii})). Thus, our method in
  Section~\ref{sec:type.I} does not apply since the measure $\mathrm
  dz$ in~\eqref{eq:fourier} is supported only on infinite-dimensional
  Hilbert spaces.  Curiously, one can construct a \emph{finitely}
  additive measure on the group dual $\hG$ supported by the set of
  finite-dimensional representations of $\hG$ (cf.~\cite{pytlik:79}).
  The group dual $\hG$ is calculated
  e.g.~in~\cite[Beispiel~1]{kaniuth:68}.
\end{example}

\begin{example}[Free groups]
\label{ex:free.group}
  Let $\Gamma = \Z^{*r}$ be the free group with $r>1$
  generators. Then $\Gamma$ is residually finite 
  (recall Example~\ref{ex:res.fin}~(ii))
  and has positive Kadison constant (cf.~\cite[Appendix]{sunada:92}). 
  Therefore, Theorem~\ref{thm:band} applies to the $\Gamma$-coverings 
  $X \to M$ specified in Section~\ref{sec:construye}.
  
  Note that $\Gamma$ is not of type~I since $\Gamma_{\mathrm
    {fcc}}=\{e\}$ (cf.~Remark~\ref{rem:type.I}~(\ref{rem:type.I.ii})).
  Such groups are called \emph{ICC (infinite conjugacy class) groups}.
  Again, for any direct integral decomposition~\eqref{eq:fourier},
  almost all Hilbert spaces $\HS(z)$ are infinite-dimensional. Finally,
  $\Gamma$ is not amenable.
\end{example}

%
%
\subsection{An example with an amenable, non-residually finite group}
%
%
\label{sec:OpenQuestion}

Kirchberg mentioned in \cite[Section~5]{kirchberg:94} an interesting
example of a finitely generated \emph{amenable} group which is not
residually finite: Denote by $S_0$ the group of permutations of $\Z$
which leave unpermuted all but a finite number of integers. We call
$A_0$ the normal subgroup of even permutations in $S_0$.  Let $\Z$ act
on $S_0$ as shift operator.  Then the semi-direct product $\Gamma: =\Z
\ltimes S_0$ is (finitely) generated by the shift $n \mapsto n+1$ and
the transposition interchanging $0$ and $1$.  Note that $\Gamma$ and
$S_0$ are ICC groups.
\begin{lemma}
  \label{S0amenable}
  The group $\Gamma$ is amenable. Moreover, $\mathfrak{R}_{\al
    F}(\Gamma)=A_0$, hence $\Gamma$ is not residually finite.
\end{lemma}
\begin{proof}
  The group $S_0$ is amenable as inductive limit of amenable groups;
  therefore, $\Gamma$ is amenable as semi-direct product of amenable
  groups (cf.~\cite[Section~4]{day:57}).
  
  The equality $\mathfrak R_{\mathcal F} (\Gamma) = A_0$ follows from
  the fact that $A_0$ is simple.
\end{proof}

\begin{proposition}
\label{prop:OpenFinRep}
Every finite-dimensional unitary representation $\rho$ of $\Gamma$
leaves $A_0$ elementwise invariant, i.e.\ $\rho(\gamma)=\1$ for all
$\gamma \in A_0$.
\end{proposition}
\begin{proof}
  Let $\al E$ be the class of countable subgroups of $\mathrm U(n)$,
  $n \in \N$, and $\al {FG}$ the class of finitely generated groups.
  Note that $\al F \subset \al E \cap \al {FG}$ and that finitely
  generated linear groups are residually finite
  (cf.~Example~\ref{ex:res.fin}~(iii)), i.e.~$\al E \cap \al {FG}
  \subset\mathfrak R \al F$.  Arguing as in the proof of
  Lemma~\ref{lem:res.class} we obtain from the inclusions $\al F
  \subset \al E \cap \al {FG} \subset\mathfrak R \al F$ that
  $\mathfrak R_{\al E \cap \al{FG}}(\Gamma) = \mathfrak R_{\al
    F}(\Gamma)$.  Now by Lemma~\ref{S0amenable} the $\al F$-residual
  of $\Gamma$ is $A_0$.  Finally, since $\Gamma$ itself is finitely
  generated (i.e.\ $\Gamma \in \al {FG}$), we have
  \begin{displaymath}
     \mathfrak R_{\al E}(\Gamma) =
     \mathfrak R_{\al E \cap \al{FG}}(\Gamma)=A_0.
  \end{displaymath}
  This concludes the proof since $\rho$ is a finite-dimensional
  unitary representation iff $\mathrm {im} (\rho) \cong \Gamma/\ker
  \rho \in \al E$, i.e.\ $\mathfrak R_{\al E}(\Gamma)$ is the
  intersection of all $\ker \rho$, where $\rho$ are the
  finite-dimensional, unitary representations of $\Gamma$.
\end{proof}

In conclusion, we cannot analyse the spectrum of $\laplacian X$ by
none of the above methods since $\Gamma$ is not residually finite (and
therefore neither of type~I). Nevertheless, equality holds
in~\eqref{eq:spec.dir.int}, but we would need infinite-dimensional
Hilbert spaces $\HS(z)$ in the direct integral decomposition in order to
describe the spectrum of the whole covering $X \to M$ and not only of
the subcovering $X/A_0 \to M$ (with covering group $\Z \times \Z_2$,
cf.\ Diagram~\eqref{eq:sub.cov}).

\begin{remark}
  Coverings with transformation groups as in the present subsection
  cannot be treated with the methods developed in this paper.  It
  seems though reasonable that even for non-residually finite groups
  the construction specified in Section~\ref{sec:construye} still
  produces at least $n$ spectral gaps, $n\in\N$. To show this one
  needs to replace the techniques of Section~\ref{sec:floquet} that
  use the min-max principle in order to prove the existence of
  spectral gaps for these types of covering manifolds.
\end{remark}

%
\section{Conclusions and applications}
%
\label{sec:outlook}

Given a Riemannian covering $(X,g)\to (M,g)$ with a residually finite
transformation group $\Gamma$ we constructed a deformed
$\Gamma$-covering $(X,g_\eps)\to (M,g_\eps)$ such that
$\spec\Delta_{(X,g_\eps)}$ has $n$ spectral gaps, $n\in\N$.
Intuitively one decouples neighbouring fundamental domains by
deforming the metric $g\to g_\eps$ in such a way that the junctions of
the fundamental domains are scaled down (cf.~Figure~\ref{fig:per-mfd}).
Therefore, our construction may serve as a model of how to use
geometry to remove unwanted frequencies or energies in certain
situations which may be relevant for technological applications.

For instance, the Laplacian on $(X,g_\eps)$ may serve to give an
approximate description of the energy operator of a quantum mechanical
particle moving along the periodic space $X$. Usually, the energy
operator contains additional potential terms coming form the curvature
of the embedding in some ambient space, cf.~\cite{froese-herbst:00},
but, nevertheless, $\laplacian {(X,g_\eps)}$ is still a good
approximation for describing properties of the particle. A spectral
gap in this context is related to the transport properties of the
particle in the periodic medium, e.g., an insulator has a large first
spectral gap.

Another application are photonic crystals, i.e.\ optical materials
that allow only certain frequencies to propagate. Usually, one has to
consider differential forms in order to describe the propagation of
classical electromagnetic waves in a medium. Nevertheless, if we
assume that the Riemannian density is related to the dielectric
constant of the material, one can use the scalar Laplacian on a
manifold as a simplified model.  For more details, we refer
to~\cite{kuchment:01,figotin-kuchment:98} and the references therein.

A further interesting line of research would be to consider the
opposite situation as in the present paper; that means the use of
geometry to prevent the appearance of spectral gaps
(cf.~\cite{friedlander:91,mazzeo:91}).  In fact, these authors proved
that $\EWN {k+1}(D) \le \EWD k(D)$ for all $k \in \N$, i.e, that $I_k
\cap I_{k+1} \ne \emptyset$ for all $k \in \N$ provided $D$ is an open
subset of $\R^n$ or a Riemannian symmetric space of non-compact type.
On such a space, we have a priory no information on the existence of
gaps. 

It would also be interesting to connect the number of gaps with
geometric quantities, e.g., isoperimetric constants or the curvature.
We want to stress that the curvature of $(X,g_\eps)$ is \emph{not}
bounded as $\eps \to 0$ (cf.~\cite{post:03a}) in contrast to the
degeneration of Riemannian metrics under curvature bounds
(cf.~e.g.~\cite{cheeger:01}).

In the present paper we have considered $\laplacian X$ as a prototype
of an elliptic operator and have avoided the use of a potential $V$.
In this way we isolate the effect of geometry on $\spec {\laplacian X}$.
Of course, our methods and results may also be extended to more
general periodic structures that have a ``reasonable'' Neumann
Laplacian as a lower bound and satisfy the spectral ``localisation''
result in Theorem~\ref{thm:spec.incl}. For example, one can also study
periodic operators like $\laplacian X + V$, operators on quantum wave
guides, more general periodic elliptic operators or operators on
metric graphs (cf.~e.g.~\cite{exner-post:05} for examples of periodic
metric graphs with spectral gaps).

Finally, we conclude mentioning that we can not apply directly our
result to disprove the Bethe-Sommerfeld conjecture on manifolds, which
says that the number of spectral gaps for a periodic operator in
dimensions $d \ge 2$ remains \emph{finite}.  Even if we know that the
spectrum of the Laplacian on $(X,g_\eps)$ converges to the discrete
set $\set{\lambda_k}{k \in \N}$ as $\eps \to 0$, we cannot expect a
\emph{uniform} control of the spectral convergence on the whole
interval $[0,\infty)$ since there are topological obstructions
(cf.~\cite{chavel-feldman:81}). Note that a uniform convergence would
immediately imply that $\spec {\laplacian{(X,g_\eps)}}$ would have an
\emph{infinite} number of spectral gaps.  Nevertheless, we hope that
our construction will contribute to the clarification of the status of
this conjecture.

%
\section*{Acknowledgements}
%

  It is a pleasure to thank Mohamed Barakat for helpful discussions on
  residually finite groups. We are also grateful to David
  Krej{\v{c}}i{\v{r}}{\'\i}k and Norbert Peyerimhoff for useful
  comments. Finally, we would like to thank Volker En{\ss},
  Christopher Fewster, Luka Grubi\v{s}i\'c and Vadim Kostrykin for
  valuable remarks and suggestions on the manuscript.




\providecommand{\bysame}{\leavevmode\hbox to3em{\hrulefill}\thinspace}

\end{document}